%% file: main.tex
\documentclass[oupdraft]{bio}
\usepackage{url}
\usepackage{xcolor,comment}

\usepackage{tabularray}
\usepackage{float}
\usepackage{lastpage}
\usepackage{graphicx}
\usepackage{codehigh}
\usepackage[normalem]{ulem}
\usepackage{algorithm}
\usepackage{algorithmic}
\UseTblrLibrary{booktabs}
\UseTblrLibrary{siunitx}

\NewTableCommand{\tinytableDefineColor}[3]{\definecolor{#1}{#2}{#3}}

\newcommand{\XX}{\mathbf{X}}
\newcommand{\YY}{\mathbf{Y}}
\newcommand{\BB}{\mathbf{B}}

\newcommand{\CC}{\mathbf{C}}
\newcommand{\indep}{\perp \!\!\! \perp}

\begin{document}

\title{Functional Moments Regression}

\author{Mingyuan Li$^1$, Martin A. Lindquist$^1$, Edward Gunning$^2$, Ciprian Crainiceanu$^{1,\ast}$\\[4pt]
\textit{$^1$Department of Biostatistics, Johns Hopkins Bloomberg School of Public Health\\$^2$School of Mathematical Sciences, University College Cork, Ireland}
\\[2pt]
{$^*$ccraini1@jhu.edu}}

\markboth%
{Li, Lindquist, Gunning, Crainiceanu}
{Functional Moments Regression}

\maketitle

\footnotetext{To whom correspondence should be addressed.}

\begin{abstract}
{The Gaussian Process (GP) assumption is often used in functional data analysis. We propose a method to assess departures from the GP assumption, both in terms of the shape of the distribution and its potential dependence on covariates, using a sequence of functional moment regressions.  Our methods are inspired by and applied to objectively measured minute-level physical activity data from the National Health and Nutrition Examination Survey (NHANES) 2011-2014 study. In this setting, we find that the GP assumption is not satisfied, quantify the associations between functional moments and covariates, and show that standard data transformations, such as the log transformation, do not resolve the discrepancy between assumptions and reality. We further show that when the effect sizes are moderate, inference on the functional fixed effects is largely unaffected by departures from the GP assumption. However, when effect sizes are  small, both inference and prediction of subject-level data can be strongly affected.  Extensive simulations support these findings. This pragmatic paper presents new methods for real data analysis, with implications for statistical methodology and for understanding human activity and health.}
Accelerometry; Functional-on-scalar regression; Higher-order moments; Non-Gaussian data;

\end{abstract}

\section{Introduction}
\label{intro}

Objectively measured physical activity using accelerometers has upended our understanding of the importance of physical activity in health research. This success stems from: (1) direct measurement of physical activity in the free-living environment; (2) the continuous, detailed, and unintrusive nature of the measurement mechanism; and (3) the long periods of time (weeks to months) that cover the heterogeneity of the full spectrum of human activities. Almost unsurprisingly, this has translated in very strong associations between these measures and health outcomes. As we noted in a recent paper \cite{Crainiceanu2026}, ``While many new health technologies have been developed in the last $20$ years, data from the simple wrist-worn accelerometer provides, by far, the strongest associations with health outcomes." Indeed, using NHANES and UK Biobank, our research group has shown that objectively measured (i.e., accelerometry-derived) physical activity is the strongest predictor of all-cause and cardiovascular mortality, outcompeting even traditional risk factors such as age, gender, and BMI \citep{Ledbetter2022,LerouxUKbio2020,LerouxNHANES2024,Smirnova2020}. We have also shown that  objectively measured physical activity is the strongest modifiable risk factor for multiple sclerosis \citep{MengMS}, Parkinson's disease \citep{ZhaoPD}, and Alzheimer's disease \citep{ZhaoAD}.  The strength of these associations is belied by the almost niche status of this area of biostatistics research.

The findings described above are based on participant-level summaries  of the accelerometry data, which compress weeks of minute-level physical activity data into a small number of subject-level summaries (e.g., mean or standard deviation). Given the strength of these associations, it makes sense to expand the search for even stronger associations that may have been ignored because of this reduction in information. It is also important to characterize the structure of the objectively measured physical activity data and its association with patient characteristics. With these goals in mind, we consider the daily minute-level activity data collected in the 2011-2014 US National Health and Nutrition
Examination Survey (NHANES) study. The study collected accelerometry data for up to seven consecutive days for tens of thousands of study participants. For simplicity, here we focus on the participant-specific minute-level average for $7{,}578$ study participants in NHANES 2011-2014; for detailed inclusion criteria, see Section~\ref{sec:results}. Therefore, the physical activity data for this paper take the form of a $7{,}578 \times 1{,}440$ dimensional matrix, where every row corresponds to a study participant and every column corresponds to a minute between midnight and midnight.

\begin{figure}
    \centering
    \includegraphics[width=\linewidth]{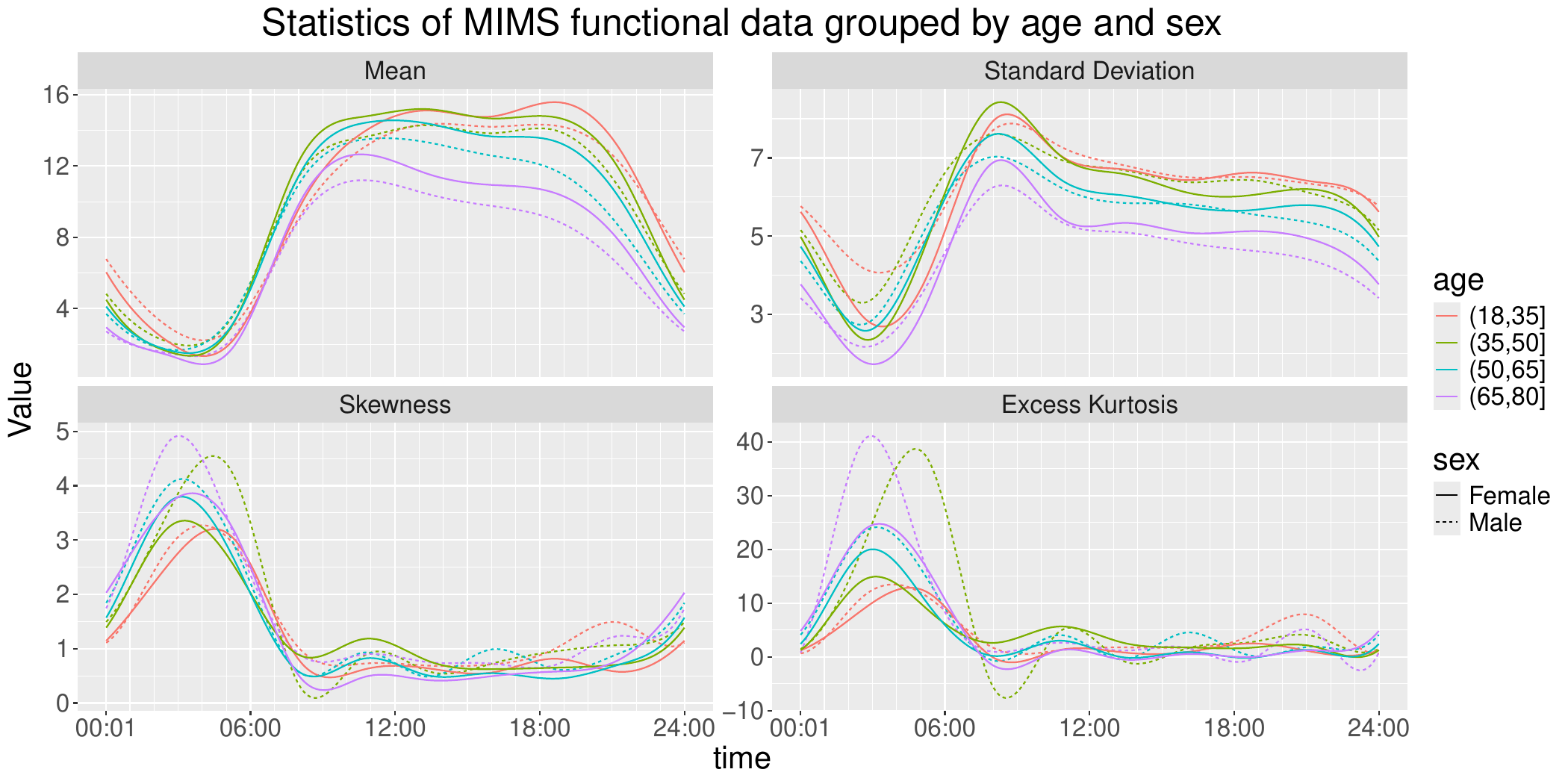}
    \caption{Smooth estimators of the mean, standard deviation, skewness, and excess kurtosis of MIMS as functions of time (x-axis). Smoothing is done using penalized cyclic cubic splines.}
    \label{fig:group-mean-original}
\end{figure}

\begin{figure}
    \centering
    \includegraphics[width=\linewidth]{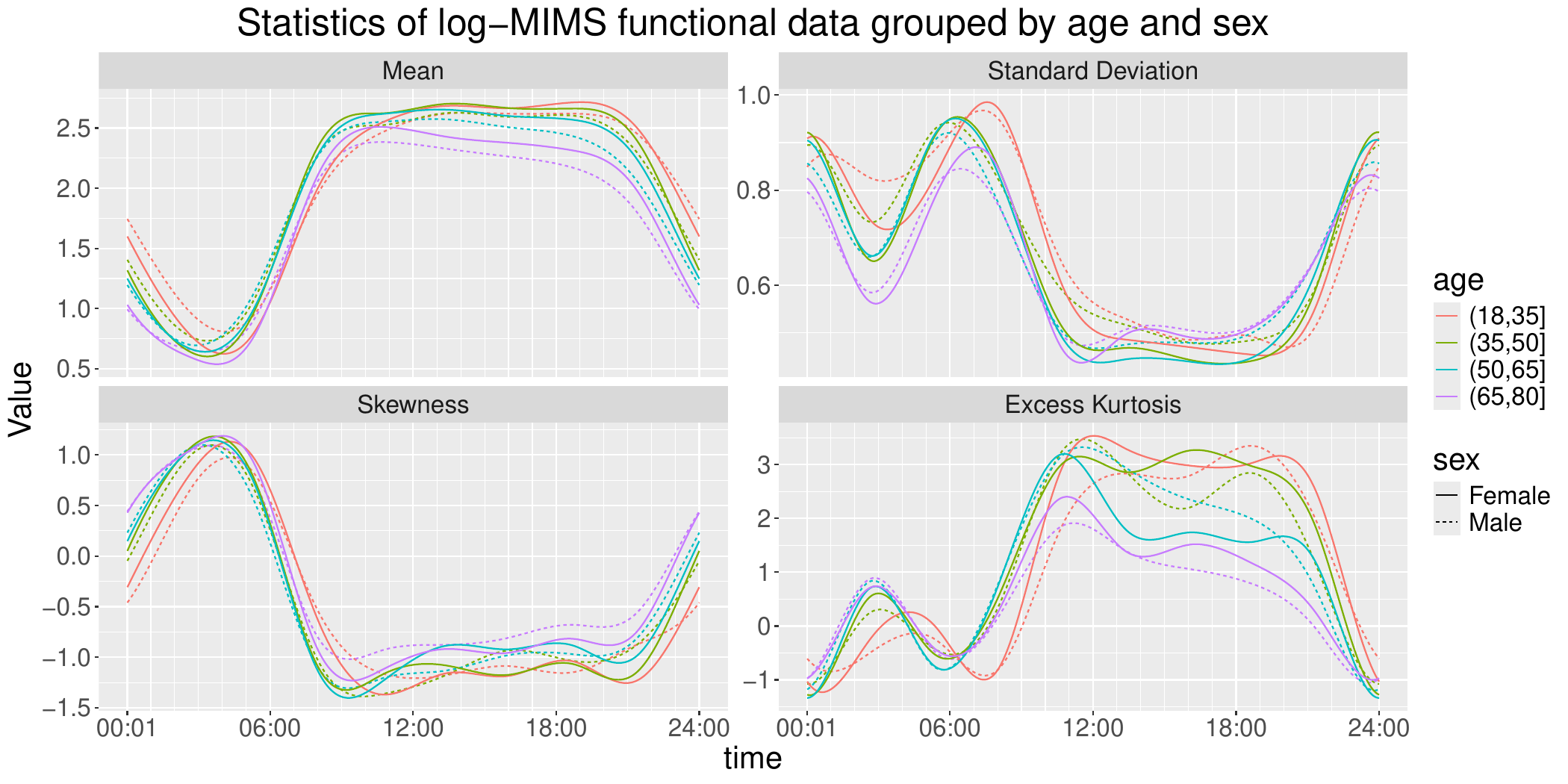}
    \caption{Smooth estimators of the mean, standard deviation, skewness, and excess kurtosis of transformed MIMS as functions of time (x-axis). Smoothing is done using penalized cyclic cubic splines.}
    \label{fig:group-mean}
\end{figure}

To better understand the underlying structure of the data, Figure \ref{fig:group-mean} displays the smooth estimates of the minute-level marginal mean, standard deviation, skewness, and excess kurtosis for the objectively measured physical activity data by age groups (color coded), with solid lines corresponding to females and  dashed lines to males. The x-axis represents time from midnight to midnight and the y-axis represents the marginal moments at each time point.
Across age and gender groups, the marginal moments exhibit substantial shape consistency and strong dependence on the time of day. 
In particular, the mean functions (top left panel) indicate that, on average within each group, individuals are more active during the day, women are more active than men within the same age group, and older individuals are less active than  younger individuals. 
The standard deviation (top right panel) curves also exhibit substantial intra-day variability, with larger values in the early morning  (around $6$ AM to $7$ AM) and lower values during the night (around $3$ AM to $4$ AM). The skewness functions (bottom left panel) indicate substantial within-day variability with higher positive skewness during the night (between $2$ AM and $4$ AM) and much smaller positive skewness during the day (around $0.5$) for all groups. 
The excess kurtosis functions (bottom right panel) are very large during the night (between $10$ and $40$) and much smaller during the day, though some excess kurtosis seems to be present in some of the subgroups.

The fact that objectively measured physical activity data do not have Gaussian marginal distributions has been noted before; see, for example, \cite{schrack2014,leroux2019,Crainiceanu:2024}. The solution proposed was to use the transformation ${\rm MIMS}\rightarrow\log(1+{\rm MIMS})$, which reduces the degree of non-Gaussianity. But by how much? Is that enough, and are there any unintended consequences? Figure~\ref{fig:group-mean} displays the same summaries as Figure~\ref{fig:group-mean-original}, but for the transformed data. A quick inspection of the skewness plot indicates that the log transformation reduced the skewness during the night, but induced negative skewness during the day, which was never reported before. Similarly, kurtosis becomes much stronger during the day, indicating that global transformations may mitigate some characteristics of the data, though they may also induce unexpected, possibly unwanted, structure into the data.

This paper is dedicated to modeling the higher-order moments of the functional data and quantifying their association with scalar covariates. To achieve this, in Section 2 we introduce a functional regression method and associated inferential tools for the moments of functional data. Section 3 provides a simulation study showing the accuracy of our method for the estimation and inference of moments of functional data. Section 4 includes the application of the method to the analysis of the NHANES MIMS data, followed by a discussion in Section 5. 

\section{Methods}
\label{method}

The data structure for each study participant $i\in\{1,2,\ldots,N\}$ is $\{\XX_i,\YY_i\}$, where $\XX_i=(X_{i1},\ldots,X_{iP})^t$ is the $P\times 1$ dimensional vector of covariates, and $\YY_i=\{Y_i(s_1),\cdots,Y_i(s_T)\}^t$ is the $T\times 1$ dimensional vector of observations at  $s_1,s_2,\ldots,s_T\in S$ within the functional domain, $S$. The complete data are stored in the $N\times P$ dimensional matrix $\XX=(\XX_1,\XX_2,\ldots,\XX_N)^t$ and the $N\times T$ dimensional matrix $\YY=(\YY_1,\YY_2,\ldots,\YY_N)^t$.

\subsection{Modeling the moments of functional data.}\label{subsec:joint_model}

With few exceptions (\citealp{Backenroth2018}), functional data models focus on modeling the mean as a function of covariates. However, in many applications, once the mean is fitted and removed, residuals exhibit substantial correlation and dependence on multiple covariates. Here we propose the following pragmatic approach, which models the first through fourth-order moments of the data and residuals.  We start by introducing models for the covariance and then expand to higher-order moments.

\subsubsection{Marginal models for the mean and covariance of functional data}\label{subsubsec:marginal_mean_cov} 

The standard function-on-scalar regression (FoSR) model (\citealp{Crainiceanu:2024, Reiss2010}) is
\begin{equation}
Y_i(s)=\sum_{p=1}^P X_{ip}\beta_p(s)+b_i(s)+\epsilon_i(s)\;,
\label{eq:basic_functional}
\end{equation}
where $\beta_p(\cdot)$ are the fixed, domain-varying functional parameters, $b_i(s)$ are independent realizations of a zero-mean random process, and $\epsilon_{i}(s)$ are independent $N(0,\sigma^2_\epsilon)$ variables, where $b_i(s)\indep \epsilon_i(r)$ for every $r,s\in S$ and $\{b_i(\cdot),\epsilon_i(\cdot)\}\indep \{b_j(\cdot),\epsilon_j(\cdot)\}$ for every $i\neq j$. Here $\beta(s,X_{i1},\ldots,X_{iP})=\sum_{p=1}^P X_{ip}\beta_p(s)$ is the marginal mean (fixed functional effect), $b_i(s)$ is the subject-specific (random functional effect), and $\epsilon_i(s)$ is the random error. 
More complex models of $\beta(s,X_{i1},\ldots,X_{iP})$ could be considered, but we use $\beta(s,X_{i1},\ldots,X_{iP})=\sum_{p=1}^P X_{ip}\beta_p(s)$  for presentation purposes.

A popular assumption is that $b_i(s)\sim GP(\mathbf{0},\mathbf{\boldsymbol{\Sigma}})$, that is $b_i(s)$ is an independent realization of a Gaussian process with zero mean and covariance $\mathbf{\boldsymbol{\Sigma}}$. This assumption is likely made because it is simple, computationally feasible, and is the direct generalization of the normality assumption to functional observations. 
However, the assumption implies that: (1) for every $s$  the residuals $r_i(s)=Y_i(s)-\sum_{p=1}^PX_{ip}\beta_p(s)=b_i(s)+\epsilon_i(s)$ are normal, symmetric around zero, and do not depend on covariates; and (2) the vectors $\{r_i(s_1),\ldots,r_i(s_T)\}^t$  are multivariate normal. These implications can  can be directly checked and, sometimes, invalidated by the estimated model residuals. The questions that we address here are: (1) whether the GP assumption is consistent with the observed data; (2) if the GP assumption is not valid, whether one can still conduct inference on the functional fixed effects $\beta_p(\cdot)$ and random effects $b_i(s)$; (3) if the GP assumption does not hold, how and what information can be recovered; and (4) whether the associated computational tools remain reasonable.

To answer these questions, we start by considering any basis $\phi_j(s)$ for $j=1,\ldots,J$ and assume that $b_i(s)$ can be expanded as
$b_i(s)=\sum\limits_{j=1}^J\xi_{ij}\phi_j(s)$,
where $\xi_{ij}$ are the corresponding scores for study participant $i$ and basis function $\phi_j(t)$. The process $b_i(s)$ is a zero mean Gaussian Process (GP) with covariance $\mathbf{\boldsymbol{\Sigma}}$ if and only if the random vector $\boldsymbol{\xi}_i=(\xi_{i1},\ldots,\xi_{iJ})^t$ has a multivariate normal distribution with zero mean and covariance $\mathbf{A}$. If $\sigma_{s_1,s_2}={\rm Cov}\{b_i(s_1),b_i(s_2)~|~ \mathbf{X}_i\}$ is the  $(s_1,s_2)$ component of $\mathbf{\boldsymbol{\Sigma}}$ and $a_{jk}$ is the $(j,k)$ entry of $\mathbf{A}$ then $\sigma_{s_1,s_2}=\sum_{j=1}^J\sum_{k=1}^Ja_{jk}\phi_j(s_1)\phi_k(s_2)$. Therefore: (1) irrespective of the choice of the basis $\phi_j(s)$ the vector $\boldsymbol{\xi}_i=(\xi_{i1},\ldots,\xi_{iJ})^t$ has a zero mean multivariate normal distribution; and (2) the covariance of the data, $\sigma_{s_1,s_2}$, depends on the locations $(s_1,s_2)$, but not on other covariates. In this paper we will use these properties to extend the model by allowing the higher-order moments of the scores to depend on covariates.

\subsubsection{Modeling the time-varying variance conditional on covariates.}\label{subsubsec:estimation_mean_cov}
We start by  modeling the marginal variance $a_{jj}=\textrm{Var}(\xi_{ij})=e^{\XX_i^t\boldsymbol{\gamma}_j}$, where $\boldsymbol{\gamma}_j$ is a $P\times 1$ dimensional vector of parameters that describe the association between the log of the variance and covariates. 
We assume for now that the correlation matrix of $\xi_{ij}$, $\mathbf{C}$, does not depend on covariates, and that $\sigma^2_\epsilon(s)=\mathrm{Var}\{\epsilon_i(s)\}$ depends on the location $s$. With these assumptions, model \eqref{eq:basic_functional} becomes
\begin{equation}
\label{eq:full_model}
    Y_i(s)=\sum_{p=1}^P X_{ip}\beta_p(s)+\sum\limits_{j=1}^J\xi_{ij}\phi_j(s)+\epsilon_i(s)\;,
\end{equation}
where the covariance of $Y_i(s)$,  $\boldsymbol{\Sigma}(s_1,s_2~|~ \mathbf{X}_i)=\mathrm{Cov}\{Y_i(s_1),Y_i(s_2)~|~ \mathbf{X}_i\}$, has the form
\[
    \boldsymbol{\Sigma}(s_1,s_2|\mathbf{X}_i)=\BB^t(s_1)\mathbf{V}(\XX_i)\BB(s_2)+\sigma^2_\epsilon(s_1)I(s_1=s_2)\;.\]
Here $\BB(s)=[\phi_{1}(s),\ldots,\phi_J(s)]^t$ is a $J\times 1$ dimensional vector, $\mathbf{V}(\XX_i)$ is the covariance matrix of $(\xi_{i1},\ldots,\xi_{iJ})$, and $I(\cdot)$ is the indicator function.
Therefore,
\begin{equation}
\label{eq:loglinear}
    \mathbf{V}(\XX_i)=\mathrm{diag}\{e^{\XX_i^t\boldsymbol{\gamma}_1/2},\ldots,e^{\XX_i^t\boldsymbol{\gamma}_J/2}\}\CC \mathrm{diag}\{e^{\XX_i^t\boldsymbol{\gamma}_1/2},\ldots,e^{\XX_i^t\boldsymbol{\gamma}_J/2}\}\;,
\end{equation}
and
\[
    \boldsymbol{\Sigma}(s_1,s_2~|~ \mathbf{X}_i)=\sum\limits_{j_1=1}^J\sum\limits_{j_2=1}^Je^{\XX_i^t\boldsymbol{\gamma}_{j_1}/2}e^{\XX_i^t\boldsymbol{\gamma}_{j_2}/2}C_{j_1,j_2}\phi_{j_1}(s_1)\phi_{j_2}(s_2)+\sigma^2_\epsilon(s_1)I(s_1=s_2)\;,\]
where $C_{j_1,j_2}$ is the $(j_1,j_2)$ entry of matrix $\CC$. The correlation coefficient can be written as:
\begin{equation}
    \rho(s_1,s_2|\mathbf{X}_i)=\frac{\boldsymbol{\Sigma}(s_1,s_2| \mathbf{X}_i)}{\sqrt{\boldsymbol{\Sigma}(s_1,s_1| \mathbf{X}_i)\boldsymbol{\Sigma}(s_2,s_2|\mathbf{X}_i)}}\;.
    \label{eq:corr_cond}\end{equation}

The fixed effects $\sum_{p=1}^P X_{ip}\beta_p(s)$ are estimated using pointwise linear regression and then applying a smoother to the resulting coefficients \citep{Crainiceanu2012, Crainiceanu:2024,Cui2021,parkstaicu2017}. The estimated residuals $\widehat{r}_i(s)=Y_i(s)-\sum_{p=1}^P X_{ip}\widehat{\beta}_p(s)$ are then calculated, where $\widehat{\beta}_p(s)$ are the estimators of $\beta_p(s)$, respectively. We then regress $\widehat{r}_i(s)$ on $\phi_j(s)$ under the independence of residuals assumption to obtain the estimators $\widehat{\xi}_{ij}$. 

To control overfitting from the large number of spline basis functions, we use penalized spline regression with a penalty on the integral of the square of the second derivative of  $b_i(s)$ \citep{osullivan1986,marxeilers1996,Wahba:1990} and restricted maximum likelihood (REML)  to estimate the penalized spline smoothing parameter \citep{ruppert2003semiparametric,Wood2011, Crainiceanu:2024}. 
We assume that $\widehat{\xi}_{ij}$ have mean zero and $\widehat{\xi}_{ij}^2\sim{\rm Gamma}(\alpha_j,~\text{rate}=\gamma_{ij})$ with mean $\mu_j=\mathrm{E}(\widehat{\xi}_{ij}^2)=\alpha_j/\gamma_{ij}$ and $\log(\mu_j)=\XX_i^t\boldsymbol{\gamma}_j$. 
The parameters $\boldsymbol{\gamma}_j$ can be estimated by regressing $\widehat{\xi}_{ij}^2$ on $\XX_i$ using quasi-Poisson GLMs. 
If $\xi_{ij}^*=e^{-\XX_i^t\boldsymbol{\gamma}_j/2}\xi_{ij}$ and $\boldsymbol{\xi}_i^*=(\xi_{i1}^*,\ldots,\xi_{iJ}^*)^t$ are the scaled scores, then
$\CC =\textrm{Var}(\boldsymbol{\xi}_i^*)$ and $\CC$ can be estimated as the empirical correlation matrix of the scaled scores $\widehat{\xi}_{ij}^*=\widehat{\xi}_{ij}e^{-\XX_i^t\widehat{\boldsymbol{\gamma}}_j/2}$. 
To obtain a smooth estimate of the noise variance, we also use penalized spline regression to estimate $\sigma^2_\epsilon(s)$. 
Specifically, the estimated noise $\widehat\epsilon_i(s)=Y_i(s)-\sum_{p=1}^P X_{ip}\widehat{\beta}_p(s)-\sum_{j=1}^J\widehat{\xi}_{ij}b_i(s)$ are squared and averaged for each time point $s$; then the averaged squared residuals $\overline\epsilon^2(s)=\frac{1}{N}\sum_{i=1}^N\widehat\epsilon_i^2(s)$ are regressed against time $s$ using penalized cyclic cubic splines, and the fitted values are used as the estimate $\widehat\sigma^2_\epsilon(s)$.

\subsubsection{Modeling the correlation conditional on the covariates.}\label{subsubsec:correlation} The standardized scores provide the basis for modeling the correlation matrix $\CC$ as a function of covariates. More precisely, let $\boldsymbol{\xi}_{i}^*\sim N\{0,\CC(\XX_i)\}$ and assume that $\CC(\XX_i)=\sum_{j=1}^Je^{\XX_i^t\boldsymbol{\gamma}_j}\mathbf{U}_j\mathbf{U}_j^t$, where $\mathbf{U}_j$ are the marginal eigenvectors \citep{parkstaicu2015} of the covariance operator of $\boldsymbol{\xi}_{i}^*$. The modeling of the eigenvalues is related to the principal regression for matrices \cite{zhaocaffo}, but with substantial differences: (1) the outcome in our approach is a set of functions, not matrices; (2) we model the eigenfunctions of the correlation matrix of functions across study participants, not the eigenfunctions of the covariance matrix within study participants; and (3) we decompose the process into an easy to implement structure.

\subsubsection{Marginal models for skewness and kurtosis.}\label{subsubsec:higher_order} So far we have worked under the hypothesis that the scores $\boldsymbol{\xi}_i$ are multivariate normal with a joint distribution that may depend on covariates. Under this assumption, the normalized scores $$\widetilde{\boldsymbol{\xi}}_i=\{\mathbf{C}(\mathbf{X}_i)\}^{-1/2}\boldsymbol{\xi}_i^*=\sum_{j=1}^Je^{-\XX_i^t\boldsymbol{\gamma}_j/2}\mathbf{U}_j\mathbf{U}_j^t\boldsymbol{\xi}_i^*\sim N(\mathbf{0}_J,\mathbf{I}_J)\;,$$ 
where $\boldsymbol{0}_J$ is the $J\times 1$ dimensional vector of zeros and $\boldsymbol{I}_J$ is the $J\times J$ dimensional identity matrix. 

However, the scores can deviate from the Gaussian assumption, or specifically, have non-zero skewness and excess kurtosis. Such deviations can occur because of the skewness and/or heavy tails of the marginal distributions of $Y_i(s)$ at any $s$.
Indeed, under our model, the third central moment of $Y_i(s)$ can be written as
\begin{equation*}
\label{eq:skewness-func}
    \begin{aligned}
        E[\{Y_i(s)-E[Y_i(s)|\mathbf{X}_i]\}^3|\mathbf{X}_i]&=\sum\limits_{j_1,j_2,j_3=1}^J\phi_{j_1}(s)\phi_{j_2}(s)\phi_{j_3}(s)E[\xi_{ij_1}\xi_{ij_2}\xi_{ij_3}]\\
        &=\sum\limits_{j_1,j_2,j_3=1}^J\phi_{j_1}(s)\phi_{j_2}(s)\phi_{j_3}(s)E[\xi^*_{ij_1}\xi^*_{ij_2}\xi^*_{ij_3}]e^{\mathbf{X}_i^t(\boldsymbol{\gamma}_{j_1}+\boldsymbol{\gamma}_{j_2}+\boldsymbol{\gamma}_{j_3})/2}\;.
    \end{aligned}
\end{equation*}
We propose to model the skewness of the data implicitly by modeling the skewness of the scores as
\begin{equation*}
E({\xi}^*_{ij_1}{\xi}^*_{ij_2}{\xi}^*_{ij_3})=\mathbf{X}_i^t\boldsymbol{\delta}_{j_1,j_2,j_3}
\label{eq:skewness_reg}
\end{equation*}
for all $j_1,j_2,j_3=1,\ldots,J$.
This implicit model for the scores corresponds to the following explicit model for the marginal third-order central moments
\begin{equation*} E[\{Y_i(s)-E[Y_i(s)|\mathbf{X}_i]\}^3|\mathbf{X}_i]=\sum\limits_{j_1,j_2,j_3=1}^J\phi_{j_1}(s)\phi_{j_2}(s)\phi_{j_3}(s)(\mathbf{X_i}^t\boldsymbol{\delta_{j_1,j_2,j_3}})e^{\mathbf{X}_i^t(\boldsymbol{\gamma}_{j_1}+\boldsymbol{\gamma}_{j_2}+\boldsymbol{\gamma}_{j_3})/2}\;.
\end{equation*}
Therefore, the marginal skewness of $Y_i(s)$ is modeled as
\[
    \text{Skewness}[Y_i(s)|\mathbf{X}_i]=\frac{\sum\limits_{j_1,j_2,j_3=1}^J\phi_{j_1}(s)\phi_{j_2}(s)\phi_{j_3}(s)(\mathbf{X}_i^t\boldsymbol{\delta_{j_1,j_2,j_3}})e^{\mathbf{X}_i^t(\boldsymbol{\gamma}_{j_1}+\boldsymbol{\gamma}_{j_2}+\boldsymbol{\gamma}_{j_3})/2}}
    {[\sum_{j_1,j_2=1}^J e^{\XX_i^t(\boldsymbol{\gamma}_{j_1}+\boldsymbol{\gamma}_{j_2})/2}\CC_{j_1,j_2}\phi_{j_1}(s_1)\phi_{j_2}(s)+\sigma^2_\epsilon(s)I(s=s)]^{3/2}}\;.
\]

For the fourth-order moments, in order to account for some cross-products being positive, we propose the following implicit models for the scores
\[E(\xi_{ij_1}^*\xi_{ij_2}^*\xi_{ij_3}^*\xi_{ij_4}^*)=\begin{cases}
e^{\mathbf{X}_i^t\boldsymbol{\eta}_{j_1,j_2,j_3,j_4}}&\sum\limits_{m=1}^4I(j_m=j)=0,2\text{ or } 4\text{ for all }j=1,2,\ldots,J\\
\mathbf{X}_i^t\boldsymbol{\eta}_{j_1,j_2,j_3,j_4}&\text{Otherwise}\;\end{cases},
\]
These implicit models correspond to explicit models of the marginal excess kurtosis for $Y_i(s)$:
\[
    \text{Kurtosis}[Y_i(s)|\mathbf{X}_i]=\frac{ E[\{Y_i(s)-E[Y_i(s)|\mathbf{X}_i]\}^4|\mathbf{X}_i]}{\boldsymbol{\Sigma}^2(s,s|\mathbf{X}_i)}\;.
\]
Notation becomes involved, and we defer the explicit formula for marginal kurtosis to the supplementary materials. All regressions and point estimators are obtained under the standard linear regression assumptions or quasi-Poisson regression for some fourth-order moments, while confidence intervals are obtained using a bootstrap of study participants. Alternative methods may include explicit modeling of errors, including the third power of a normal distribution for the skewness and the fourth power of the Normal for the kurtosis models; we do not pursue these approaches here.

\subsection{Correlation and multiplicity adjusted confidence and prediction intervals}\label{subsubsec:GCMA}

We focus now on estimating: (1) the variability of the estimators of functional parameters, $\beta_p(s)$, which can be used to construct confidence intervals; and (2) the total residual variance $\text{Var}\{b_i(s)+\epsilon_i(s)|\mathbf{X}_i\}$, which can be used to construct prediction intervals. The former approach was described in \cite{Crainiceanu:2024} in the context of symmetric distributions of functional estimators. We will extend their approach to the case in which this assumption does not hold for the functional estimators, which can happen when functional observations exhibit strong non-Gaussianity and the sample size is not sufficient to ensure that the Central Limit Theorem applies. We also discuss prediction intervals and, for the first time, we provide prediction intervals for non-Gaussian functional outcomes.

Suppose that we are interested in conducting inference on a functional parameter denoted by $g(s)$.
Inferential procedures typically involve constructing confidence bands for the parameter function $g(\cdot)$ at the observed time points $\mathbf{g}=\{g(s_1),g(s_2),\cdots,g(s_T)\}$, based on a point estimator $\widehat{\mathbf{g}}$.
\cite{Crainiceanu:2024} suggested building confidence intervals of the type
$\widehat{\mathbf{g}}\pm q_\alpha{\rm diag}\{\widehat{\text{Var}}(\widehat{\mathbf{g}})\}^{1/2}$, 
where $\widehat{\text{Var}}(\widehat{\mathbf{g}})$ is an estimator of $\text{Var}(\widehat{\mathbf{g}})$, ${\rm diag}(\mathbf{A})$ represents the diagonal vector of the matrix $\mathbf{A}$, and $v^{1/2}$ represents the component-wise square root of the vector $v$.
The quantile $q_\alpha$ can be calculated either from the multivariate quantile of $N(\mathbf{0},\widehat C_\mathbf{g})$, where $\widehat C_\mathbf{g}$ is the estimated correlation matrix of $\widehat{\mathbf{g}}$, or from bootstrapped estimates of the maximum absolute statistic~\citep{Crainiceanu:2024}:
\begin{equation*}
	D=\max\limits_{t=1,2,\cdots,T}\left|\frac{\widehat g(s_t)-g(s_t)}{ \text{Var}\{\widehat g(s_t)\}}\right|\;.
\end{equation*}
The first approach relies on the assumption that the distribution of $\widehat{\mathbf{g}}$ is Gaussian, while the second relies on the assumption that the distribution is symmetric. These assumptions may not hold for smaller sample sizes when observed data are non-Gaussian and when the smoothing parameter is close to the boundary \citep{lme_test,spline_test}. Indeed, when this happens in repeated samples, functional estimators become a mix of linear and non-linear functions that produce highly asymmetric pointwise confidence intervals.

Building confidence intervals while assuming symmetry and normality of the functional effects estimators is based on the assumption that the multivariate CLT can be applicable. When not enough observations are available to overcome the non-Gaussianity of the observed data, there is a need for alternative approaches that produce non-symmetric confidence intervals. In the case of subject-level prediction, the assumption of normality of $b_i(s)+\epsilon_i(s)$ may not hold. In this case there is no CLT to reduce the severity of the problem. In the subject-level prediction context, the prediction intervals for  $Y_i(s)$ are directly impacted by the distribution of the data. Therefore, if the observed data are asymmetric and/or have heavier tails, approaches for prediction need to take these departures from Gaussianity into account.  This is highly relevant in the NHANES physical activity dataset, which motivates this paper. However, we have found similar problems in other datasets, including for Continuous Glucose Monitoring (CGM) \citep{gaynanova2020,sergazinov2022case} and Diffusion Tensor Imaging (DTI) \citep{staicu2012}. 

To address this problem, we propose to extend the bootstrap method of \cite{Crainiceanu:2024} to allow for asymmetric confidence and prediction bands. The idea is to replace the $\alpha$-quantile $q_\alpha$ with two different quantiles, $q_{L,\alpha}$ and $q_{U,\alpha}$ and construct confidence intervals of the type
\[
	[\widehat{\mathbf{g}}-q_{L,\alpha}\;{\rm diag}\{\widehat{\text{Var}}(\widehat{\mathbf{g}})\}^{1/2},\widehat{\mathbf{g}}+q_{U,\alpha}\;{\rm diag}\{\widehat{\text{Var}}(\widehat{\mathbf{g}})\}^{1/2}]\;.
\]
When $q_{L,\alpha}=q_{U,\alpha}$, this confidence band reverts to the symmetric one introduced by \cite{Crainiceanu:2024}.
This confidence interval has a $1-\alpha$ joint coverage probability if 
\begin{equation*}
	P(q_{U,\alpha}\le\frac{\widehat g(s)-g(s)}{\widehat{\text{Var}}\{\widehat g(s)\}^{1/2}}\le q_{L,\alpha} \text{ for all\;} s=s_1,s_2,\cdots,s_T )= 1-\alpha
\end{equation*}
If we define $Z_{\max}=\max\limits_{t=1,2,\cdots,T}\frac{\widehat g(s_t)-g(s_t)}{\widehat{\text{Var}}\{\widehat g(s_t)\}^{1/2}}$ and $Z_{\min}=\min\limits_{t=1,2,\cdots,T}\frac{\widehat g(s_t)-g(s_t)}{\widehat{\text{Var}}\{\widehat g(s_t)\}^{1/2}}$, this probability statement is equivalent to
$P(Z_{\max}>-q_L~\text{or}~Z_{\min}<-q_U)\le \alpha$.
This is satisfied when
\begin{equation*}
	P(Z_{\max}>-q_{L,\alpha})=P(Z_{\min}<-q_{U,\alpha})=\alpha/2\;.
\end{equation*}
Therefore, it is sufficient to find the $1-\alpha/2$ and $\alpha/2$ quantiles of $Z_{\max}$ and $Z_{\min}$, respectively. 
Suppose that we have $B$ bootstrap samples $\widehat{\mathbf{g}}_1^*,\widehat{\mathbf{g}}_2^*,\ldots, \widehat{\mathbf{g}}_B^*$ from the distribution of $\widehat{\mathbf{g}}$ where $\widehat{\mathbf{g}}_b^*=\{g_b^*(s_1),g_b^*(s_2),\cdots,g_b^*(s_T)\}$, and let
$
	\bar{\mathbf{g}}=\{\bar g(s_1),\bar g(s_2),\cdots,\bar g(s_T)\}=\frac{1}{B}\sum\limits_{b=1}^B \widehat{\mathbf{g}}_b^*$.
The bootstrap estimate of the ${\rm diag}\{\text{Var}(\widehat{\mathbf{g}})\}$ is
$\widehat{\boldsymbol{d}}=\frac{1}{B}\sum_{b=1}^B {\rm diag}\{(\widehat{\mathbf{g}}_b^*-\bar{\mathbf{g}})(\widehat{\mathbf{g}}_b^*-\bar{\mathbf{g}})^t\}$.
Let $\widehat d_t$ be the $t$-th entry of $\widehat{\boldsymbol{d}}$, then the $b$ bootstrap samples of $Z_{\max}$ and $Z_{\min}$ are
\begin{equation*}
	Z_{\max, b}^*=\max\limits_{t=1,2,\cdots,T}\frac{g_b^*(s_t)-\bar g(s_t)}{d_t^{1/2}},\;Z_{\min, b}^*=\min\limits_{t=1,2,\cdots,T}\frac{g_b^*(s_t)-\bar g(s_t)}{d_t^{1/2}}\;.
\end{equation*}
Then an asymmetric $1-\alpha$ confidence band is
\begin{equation}
	\label{eq:global-asym-CI-full}
	[\widehat{\mathbf{g}}-q_{L,\alpha}\;{\rm diag}\{\text{Var}(\widehat{\mathbf{g}})\}^{1/2},\widehat{\mathbf{g}}+q_{U,\alpha}\;{\rm diag}\{\text{Var}(\widehat{\mathbf{g}})\}^{1/2}]\;,
\end{equation}
where $q_{U,\alpha}$ is the $1-\alpha/2$ quantile of $-Z_{\min,1}^*,\ldots,-Z_{\min,B}^*$, and $q_{L,\alpha}$ is the $\alpha/2$ quantile of $-Z_{\max,1}^*,\ldots,-Z_{\max,B}^*$.
These confidence bands retain some information on the skewness of $\widehat{\mathbf{g}}$ by using different upper and lower quantiles, $q_{L,\alpha}\neq q_{U,\alpha}$, though they assume that the same level of asymmetry holds across the functional domain $s=s_1,\ldots,s_T$.
This assumption may not hold in cases when functional observations  are left-skewed in some sub-domains and right-skewed in another, or when the level of skewness varies substantially along the functional domain.

\section{Simulations}
\label{sec:simulation}

We evaluate the empirical performance of our stepwise regression method for estimating the fixed effects and the conditional moments.
We set the number of fixed effects to $p=3$ and
the functional responses were simulated on an equally spaced grid $\{s_k=\frac{k}{K}:k=1,2,\ldots,K\}$ on $S\in[0,1]$. 
For subject $i$ at time $s$, the data-generating model was
\[
Y_i(s)=\sum\limits_{l=1}^4\beta_l(s)X_{il}+\sum\limits_{j=1}^5\xi_{ij}\phi_j(s)+\epsilon_i(s)\;,
\]
which contains $L=4$ fixed effects and $J=5$ random effects. We further assume that
$\epsilon_i(s)$ are independent $N(0,\sigma_\epsilon^2(s))$ and
$(\xi_{i1},\ldots,\xi_{i5})$ is a mean-zero $J=5$-dimensional random vector independent of $\epsilon_i(s)$. The basis functions $\phi_j(s)$, $j=1,\ldots,5$ are periodic $3^{\textrm{rd}}$-degree B-splines  with knots at $\{0,\frac{1}{5},\frac{2}{5}\ldots,1\}$, respectively.
The functional fixed effects and noise variance parameters were chosen to emulate as closely as possible the estimated structure of the data analyzed in Section~\ref{sec:results}. 
The noise variance $\sigma_\epsilon^2(s)$ was set to
\(
\sigma_\epsilon^2(s)=0.1+0.35\exp\{-4\{1-\cos[(2s-0.5)\pi]\}\}+0.25\exp\{-4[1-\cos(2s\pi)]\}
\). 
To induce skewness and kurtosis in the functional response, we generate random effect coefficients by generating Gaussian random vectors and applying inverse-quantile transformations to them; precise details on the fixed effects, noise variance, and random effect generation are provided in the Supplementary Material.

Data were generated for $N=100,~300~,1000$ study participants and  $K=1440$ , $480$, $144$ observations per study participant, respectively.
The choice of number of observations was set to emulate daily objectively measured physical activity data with measurement frequencies of $1$, $3$, and $10$ minutes, respectively. 
For each setting, we conducted $I=100$ simulations. 
The covariates were independently sampled with $X_{i1}=1$, $X_{i2}\sim \text{Uniform}(-30,30)$, $X_{i3}\sim \text{Bernoulli}(0.5)$, $X_{i4}\sim \text{Bernoulli}(0.3)$.
Models were fit using the same spline basis as in Section~\ref{subsec:analysis-result} and  $B=100$ bootstrap samples were used for inference.

The estimation accuracy for all functional parameters was quantified by the integrated squared error (ISE) conditional on the covariates $(1,-10,0,0)$ and $(1,10,0,0)$. 
For a functional parameter $f(s)$ and an estimator $\widehat{f}(s)$, we define $\mathrm{ISE}(\widehat{f})=\int_0^1\{\widehat{f}(s)-f(s)\}^2\mathrm{d}s$.
The performance of the confidence intervals was quantified by the empirical coverage probability of $95$\% pointwise and CMA confidence bands.
For pointwise confidence intervals, the average coverage frequency over all locations was calculated. 
For simultaneous confidence intervals, the average coverage frequency at all locations simultaneously was calculated.

\subsection{Simulation Results}\label{subsec:sim_results}
\begin{figure}
    \centering
    \includegraphics[width=\linewidth, page=1]{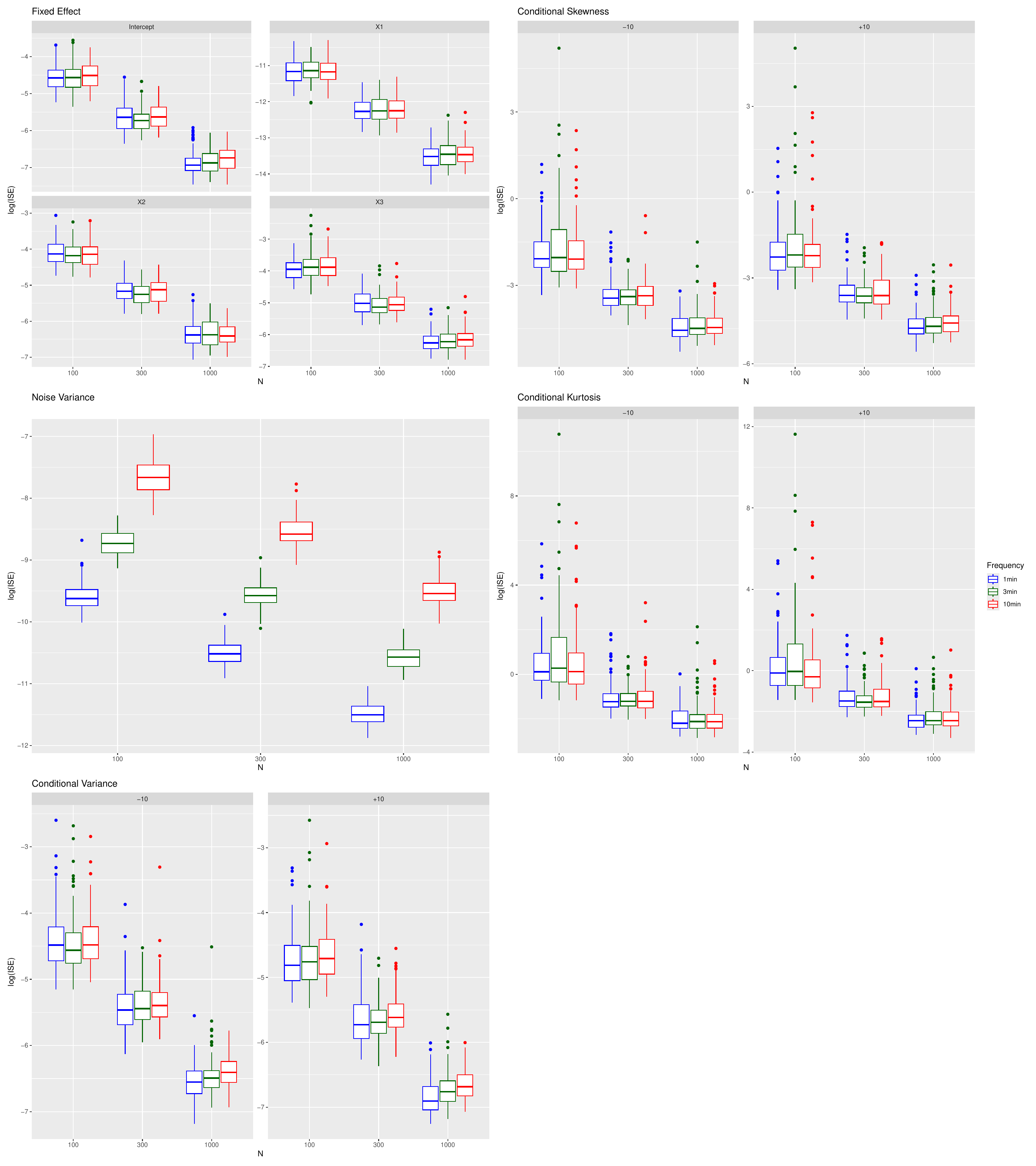}
    \caption{Boxplot of 100 simulated integrated square error (ISE) of multiple functional parameters, under different number of individuals (N) and different sampling frequencies. 
    First column top to bottom: fixed effects, noise variance, conditional variance, conditional skewness, conditional kurtosis. For fixed effects, the subpanels from left to right, from top to bottom are the intercept and the three simulated covariates. For conditional statistics, the subpanels on the left and the right indicate covariate values of $(-10,0,0)$ and $(10,0,0)$. The ISE are plotted in log scale.}
    \label{fig:ise}
\end{figure}

Figure \ref{fig:ise} shows the boxplots of integrated squared error (ISE) for all functional estimators separated by the number of observations (blue = every $1$ minute, green = every $3$ minutes, and red = every $10$ minutes). Boxplots are also grouped by the number of observations (x-axis) from left to right. ISE for all parameters decreases with the increased number of study participants, $N$. With the exception of the noise variance, ISE does not seem to vary substantially with the number of observations per study participant in the range considered. This is likely due to the fact that the variation in the functional model parameters is very smooth across the domain. The ISE of the noise variance depends strongly on both the number of subjects and the number of observations per subject. This is due to the fact that the variance increases quickly between sleep and non-sleep periods, and this is better captured by having more observations in the areas where variance changes rapidly. It is also worth noting that the log ISE for variance, skewness, and kurtosis exhibits many outliers, corresponding  to heavier tails.

Tables \ref{tab:coverage_FE_noise} and \ref{tab:coverage_cond_moments} summarize the performance of confidence intervals for the functional parameters; Figure \ref{fig:coverage} in the supplementary materials displays the same information. 
For all functional parameters, the Wald intervals have average coverage probability close to $95$\%, while the CMA intervals tend to have lower coverage than $95$\%, especially for the conditional variance parameters. 
The asymmetric CMA confidence intervals do not necessarily have better coverage than the symmetric CMA confidence intervals. 
This is likely due to the fact that the pointwise distributions of the data and, in turn, the estimators, have different directions of asymmetry along the time domain (strong positive skewness during the night and slight negative skewness during the day). This could lead to uneven coverage probability of the symmetric Wald intervals while retaining average coverage close to the nominal $95$\%.

\section{Results}
\label{sec:results}

\subsection{Data Description}
\label{subsec:datadesc}

The National Health and Nutrition Examination Survey (NHANES) is a large ongoing study conducted by the Centers for Disease Control and Prevention (CDC) of the United States to assess the health and nutritional status of the US population. 
Here we focus on the data from the 2011-2012 and 2013-2014 NHANES waves, which include objectively measured physical activity data in the free-living environment. 
Each participant was instructed to wear a wrist-worn tri-axial accelerometer for seven consecutive days. 
The acceleration was collected at $80$Hz, and the data were summarized by NHANES in minute-level Monitor Independent Movement Summary (MIMS) units \citep{john2019open}. 
For each study participant, data were averaged at the same minute across valid days and only individuals who had at least three valid days as defined by \citep{Crainiceanu:2024,LerouxNHANES2024} were retained for analysis. Therefore, for each study participant the objectively measured physical activity was summarized by a $1{,}440$ dimensional vector, where each entry corresponds to one minute of the day from midnight to midnight. 
For the purpose of this manuscript, the data were further transformed at the minute level using the transformation $x\to \log(1+x)$, which preserves the ranking of the observations, reduces the skewness of the marginal distributions, and transforms zero into zero. 
This transformation was first suggested by \cite{schrack2014} for activity counts obtained from hip-worn accelerometers in the NHANES 2003-2006 study \citep{leroux2019}. 
The MIMS values obtained from wrist-worn accelerometers in the NHANES 2011-2014 \citep{Crainiceanu:2024}  are far less skewed than the activity counts, but the transformation is still considered here, as it is easy to use and reasonable. 
In addition to minute-level physical activity data, we consider the following covariates: age, gender, race, body mass index (BMI), education level, poverty income ratio (PIR), and history of coronary heart disease (CHD). 
These covariates were chosen because they are thought to be associated with physical activity and for consistency with previous studies. 
We exclude participants with incomplete covariates and those with ``Refused'' or ``Don't know'' recorded in education and CHD variables, and include $7{,}578$ participants in our analysis. Table \ref{tab:tab1-char} provides summaries of the covariates for this subset of the NHANES 2011-2014 data.

\input{tables/table_1_characteristics}

Therefore, for each of the $7{,}578$ participants, the data consist of minute-level summaries ($1{,}440$ observations per day) of accelerometer-derived physical activity paired with the covariates summarized in  Table~\ref{tab:tab1-char}.
Figure \ref{fig:group-mean} displays the smooth estimates of the marginal mean, standard deviation, skewness, and excess kurtosis for the objectively measured physical activity data by age groups (color coded) and gender (solid lines correspond to females and  dashed lines to males). The x-axis is time from midnight to midnight, and the y-axis shows the marginal moments at each time point.
To account for the periodic nature of physical activity,  cyclic penalized splines were used to smooth the data.

Across age and gender groups, the marginal moments exhibit substantial shape consistency and strong dependence on the time of day. 
In particular, the mean functions (top left panel) indicate that, on average within each group, individuals are more active during the day, women are more active than men within the same age group, and older individuals are less active than  younger individuals. 
The standard deviation (top right panel) curves also exhibit substantial intra-day variability, with larger values around midnight and in the early morning hours (standard deviations around $0.8$ to $0.9$) and lower values during the day (standard deviations around $0.5$). 
This may be counter-intuitive, though it may be due to the log transformation of the MIMS values, whose means are much larger during the day on the original scale. 
The skewness functions (bottom left panel) also indicate substantial within-day variability with higher positive skewness during the night ($0.5$ to $1$) and negative skewness during the day (around $-1$) for all groups. 
The negative skewness during the day may be due to ``over-transforming'' the MIMS values during the day by taking the log transformation. 
The excess kurtosis functions (bottom right panel) are smaller during the night with large values during the day, especially for individuals between $18$ and $35$ (red lines) and $35$ and $50$ (green lines).

One could argue that the log data transformation induced these problems, which may not exist on the original scale. 
Alas, this is real-world data, and the problems on the original scale are even more pronounced. 
Indeed, Figure~\ref{fig:group-mean-original} displays the same time-varying moments estimated on the original (not log-transformed) MIMS data. 
The units are changed (different y-axes), but the overall story is consistent. 
All moments exhibit substantial within-day variation and strong (at least visual, if not yet inferential) dependence on age and gender. 
The means follow the same patterns as the means on the log-scale, indicating lower activity during the night and higher activity during the day. 
The standard deviation of the MIMS data is larger during the day, whereas it was smaller for the log-MIMS data. 
The skewness and kurtosis of the MIMS data are very large during the night and much smaller, though still consistent with non-Gaussian distributions, during the day.

Once moments of the functional data are summarized as in Figures~\ref{fig:group-mean-original} and \ref{fig:group-mean}, it becomes clear that Gaussian Processes (GPs) could not capture: 
(1) the potential dependence on the covariates; and 
(2) the skewness and kurtosis of the observed functional data across the entire domain. 
This raises questions about how much information may actually be lost by using an inappropriate Gaussian Process assumption and what, if any, are the downstream implications of such a modeling choice are. 

\subsection{Analysis Results}
\label{subsec:analysis-result}

To quantify and test the dependence of the first four moments on time and covariates, we  employ the sequential modeling strategy described in Section~\ref{subsubsec:estimation_mean_cov} using the covariates introduced in Section~\ref{subsec:datadesc}. Because physical activity is periodic, we use penalized cyclic cubic spline regression to model the time-varying fixed effects. 
Confidence bands are generated via nonparametric bootstrap using $B=1,000$ resampled datasets. 
To preserve interpretation when modeling the covariance, we use the  cyclic cubic spline basis, $\phi_j(s)$, for $j=1,\ldots,J=24)$ with equidistant internal knots at 1:00, 2:00, $\ldots$, 23:00 and boundary knots at 0:00 and 24:00. 

\subsubsection{Estimation of the fixed effects components of the mean.}\label{subsubsec:mean}

\begin{figure}
	\centering
	\includegraphics[width=0.9\linewidth]{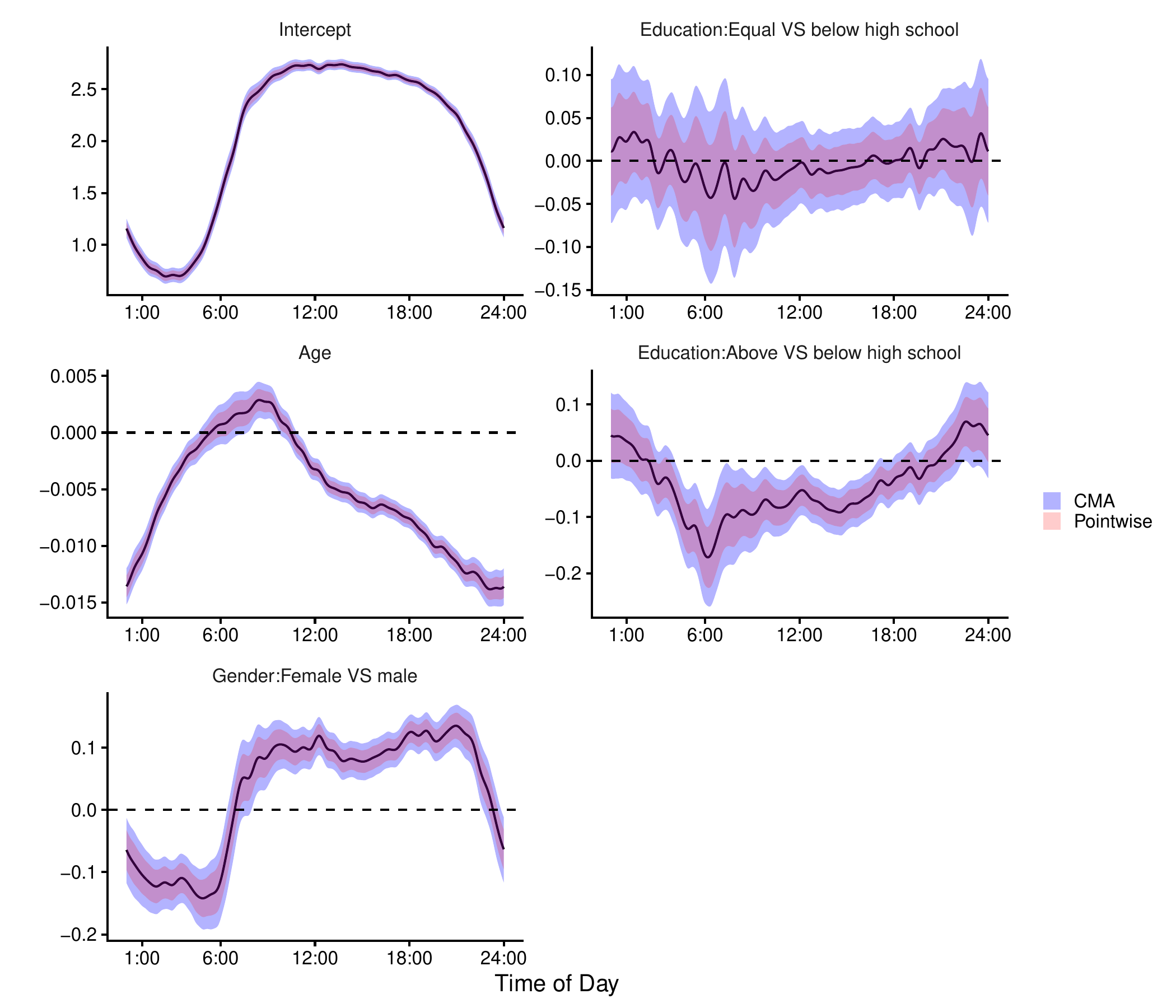}
	\caption{Estimated fixed effects (black solid lines) of subject characteristics on logged MIMS data in NHANES study. Pointwise (blue) and Correlation and Multiplicity Adjusted (CMA, red) 95\% confidence bands. Continuous predictors such as age, BMI and poverty income ratio are centered. From left to right, top to bottom: intercept term, age (in years), gender (female), education: high school equivalent, and more than high school.}
	\label{fig:fixed-effect-estimate}
\end{figure}

Figure \ref{fig:fixed-effect-estimate} displays the estimated intercept and fixed effects of age, gender, and education on the mean structure, along with the $95$\% pointwise  and Correlation and Multiplicity Adjusted (CMA) confidence bands \citep{Crainiceanu:2024}.
The estimated intercept, shown in the top left panel, represents the expected log-MIMS function for an individual who is Mexican-Hispanic, has no CHD history, lower-than-high-school education (reference groups for race, CHD, and education variables) and average age, BMI, and PIR.
The shape of the intercept estimate is consistent with previous functional analyses of  the NHANES physical activity data \citep{Cui2021, Crainiceanu:2024,LerouxNHANES2024}. Therefore, on average, the activity for individuals with these covariates  is lowest during the night, increases from early morning until about midday, stays high and relatively constant throughout the afternoon and early evening, and steadily decreases later in the evening.
The estimated effect of age, shown in the middle left panel, indicates that age has a statistically significant association with physical activity during the entire day except around 6 AM and 11 AM. Specifically, older participants are less active in the afternoon and during the night, and more active in the morning (7 AM to 11 AM).
The bottom left panel indicates that females are less active during the night than males, but more active during the day and evening.
The effect of education, displayed in the two panels on the right, suggests that compared to those with education lower than high school, having high-school-equivalent education has no statistically significant effect on physical activity at any time during the day. However, individuals with education above high school  tend to be significantly less active during the day but not during the night.

\subsubsection{Modeling the variance and correlation.}\label{subsubsec:variance_corr}

\begin{figure}
    \centering
    \includegraphics[width=.8\linewidth]{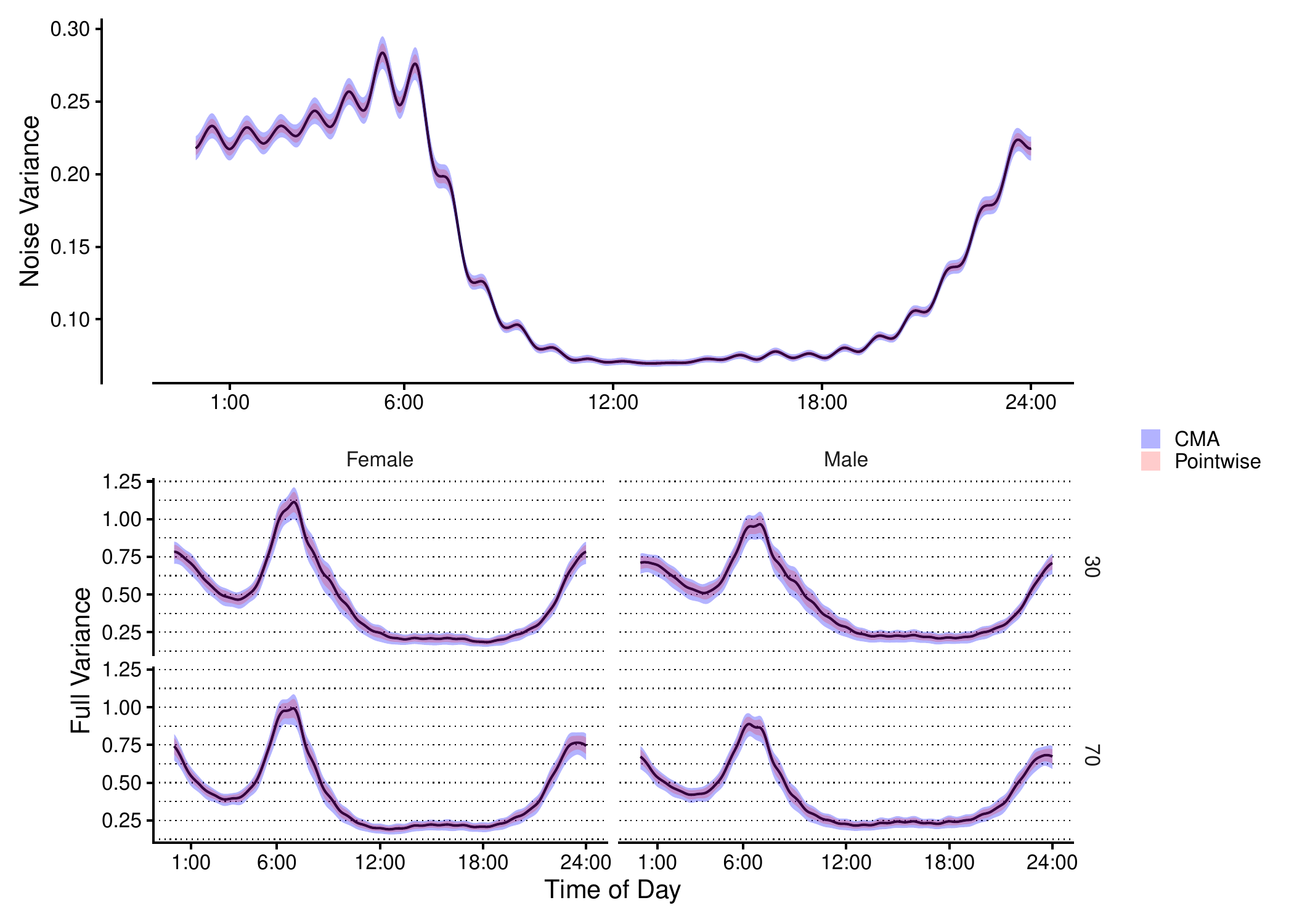}
    \caption{Estimated noise variance after accounting for time-varying fixed effects of the mean (top panel) and full residual variance (bottom panel) of the log-transformed MIMS outcome. Male and female individuals aged 30 and 70 who are non-hispanic white, have education above high school level, have no history of coronary heart disease, BMI of 21 and poverty index ratio (PIR) of 2.5 are presented. The noise variance estimator is smoothed using penalized cyclic cubic splines. Pointwise confidence intervals (blue) and Correlation and Multiplicity Adjusted (CMA, red) 95\% confidence bands are included.}
    \label{fig:resid-var}
\end{figure}

\begin{figure}
    \centering
    \includegraphics[width=\linewidth]{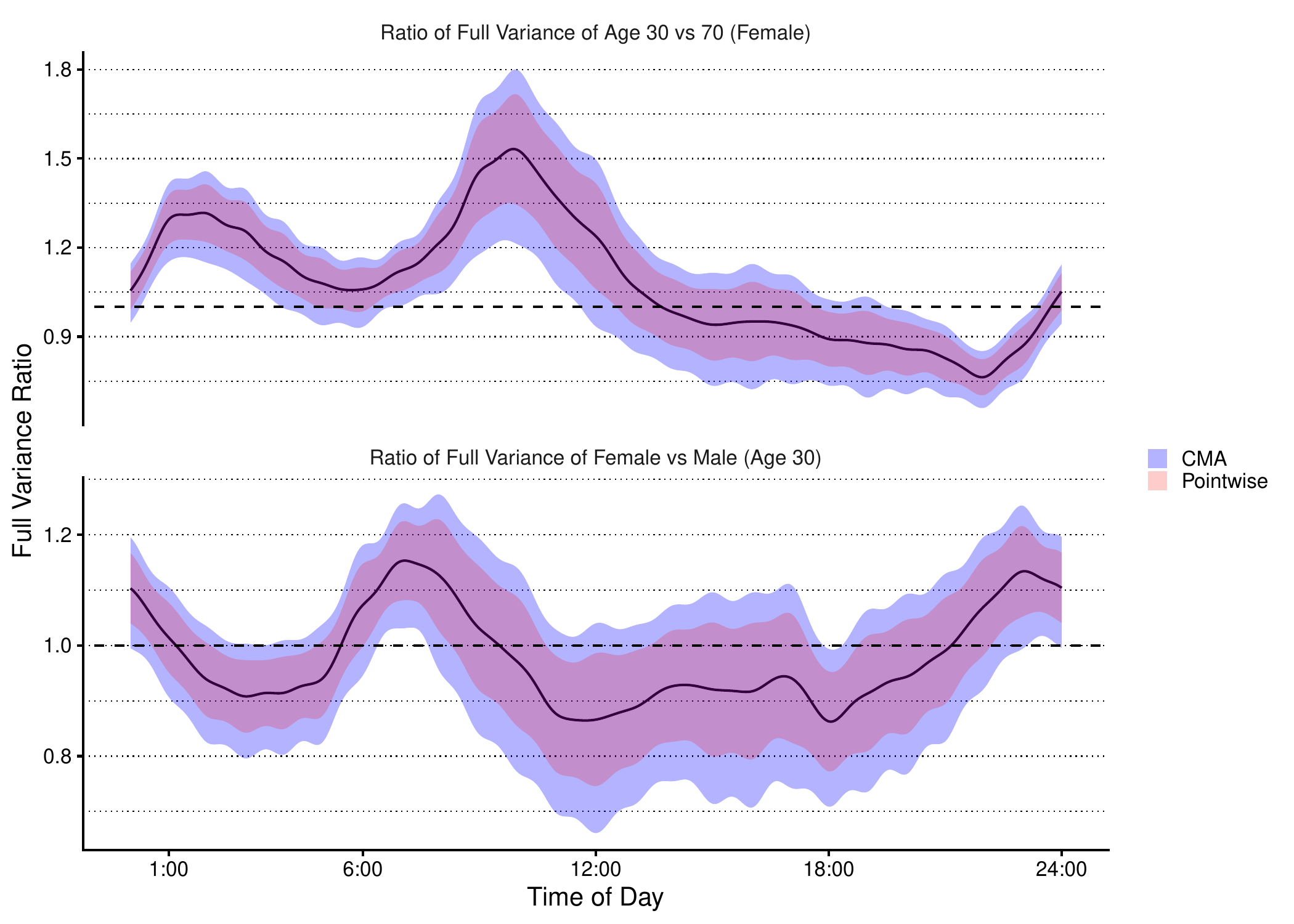}
    \caption{Estimated ratio of residual variance between females aged 30 and 70 (top panel) and between female and males aged 30 (bottom panel). Both panels correspond to individuals who are non-hispanic white, have education above high school level, no history of coronary heart disease, BMI of 21 and poverty index ratio (PIR) of 2.5;  Pointwise confidence intervals (blue) and asymmetric Correlation and Multiplicity Adjusted (CMA, red) 95\% confidence bands are included.}
    \label{fig:resid-var-ratio}
\end{figure}

The top panel of Figure~\ref{fig:resid-var} displays the estimated variance function $\widehat{\sigma}^2_\epsilon(s)$ for the log-MIMS residuals   after removing the covariate-dependent estimator of the mean.
The shape implies that the variance is not constant as a function of time. The  variance is highest around 6 AM, decreases quickly  between 6 AM and 12 PM, remains relatively small until 6 PM, and increases again quickly between 7PM and 12AM.  The lower panels in Figure \ref{fig:resid-var} display the model-based estimates of the time-varying conditional variance, $\widehat{\text{Var}}\{Y_i(s)|X_i\}$, as a function of gender and age (comparing estimated variance for $30$ and $70$ year-old males and females, respectively). The general shape of the  variance remains consistent across gender and age groups, though the magnitude of the variance changes considerably between groups. For example, males tend to have smaller variance than females both at age $30$ and $70$, with differences being most pronounced in the early morning. To see this, compare the left (corresponding to females) and right (corresponding to males) panels in the second and third rows, respectively. 
Similarly, older individuals have less variability  at most times of the day, with the largest reduction in variability during the morning hours. 
To see this, compare the second (corresponding to age $30$) and third (corresponding to age $70$) rows for both females and males.
To closely examine the magnitude of the difference between subgroups, we also conduct estimation and inference on the ratio of the residual variances. The top panel in Figure \ref{fig:resid-var-ratio} provides an estimator of the variance ratio between individuals who are $30$ and $70$ years old over the day. Just as in the case of the mean function, we display both the pointwise and CMA confidence intervals, which is likely the first time this was done on the higher-order moments of functional data. We conclude that the PA variance is $20$-$40$\% higher for $30$-year  compared to $70$-year old individuals between 12 AM and 4 AM, up to $50$\% higher between 11 AM and  1 PM, and up to $20$\% lower around 10 PM.
The bottom panel in Figure \ref{fig:resid-var-ratio} is similar to the top panel but displays the variance ratio between females and males. We conclude that the PA variance is slightly higher ($10$ to $20$\%) for females between 8 AM and 10 AM and slightly lower between 6 PM and 7 PM.
Results also indicate that the residual variance (after removing the fixed effects) varies substantially with time and among covariate groups. These differences are large and highly statistically significant.

\begin{figure}
    \centering
    \includegraphics[width=.75\linewidth]{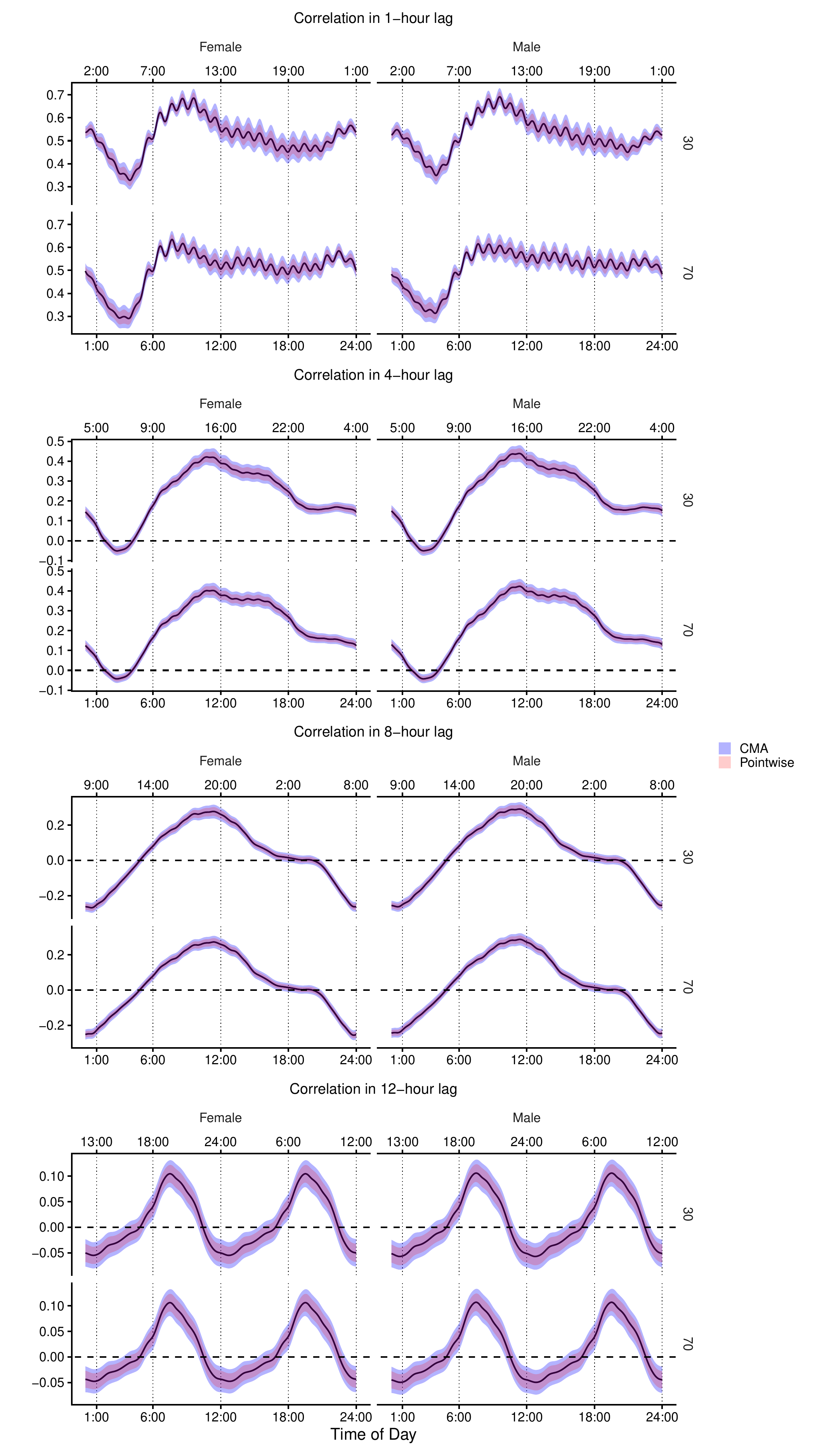}
    \caption{Model-based estimated correlation in 1, 4, 8 and 12-hour lag (1st to 4th panel) of log-MIMS. Sub-panels correspond to males and females aged 30 and 70, who are non-hispanic white, have education above high school level,  no history of coronary heart disease, BMI of 21 and poverty index ratio (PIR) of 2.5. Pointwise (blue) and asymmetric Correlation and Multiplicity Adjusted (CMA, red) 95\% confidence bands are included.}
    \label{fig:corlag-panels}
\end{figure}

Using equation~\eqref{eq:corr_cond} we estimate the covariate-dependent correlations $\widehat\rho(s,s+L|\mathbf{X})$ for the same four subgroups described above. Figure \ref{fig:corlag-panels} displays the correlation  between observations at a particular time of the day, $s$, and future time of the day, $s+T$. Time wraps around at midnight, so that $T=4$ hours into the future of $s=11$ PM is $s+T=3$ AM; this is a slight abuse of notation, but respects the actual clock time. The first through the fourth panels in Figure \ref{fig:corlag-panels} display the correlation coefficient estimates for every time point $s$ during the day at lag $T= 1, 4, 8, 12$ hours, respectively.  Just as with the other moments, we display the pointwise and CMA confidence intervals. We opted for multiple panels corresponding to a discrete set of $T$ for interpretability purposes. However, these plots could be shown as a continuous bivariate function of $s$ and $T$.

The first panel (corresponding to lag $T=1$ hour correlation)  displays positive correlations throughout the day, indicating that physical activity at every time during the day is positively associated with physical activity one hour later. However, correlations vary substantially during the day from around $0.4$ during the early morning hours ($3$ AM to $5$ AM), to $0.5$ to $0.7$ during daytime ($8$ AM to $6$ PM) and  evening ($6$ PM to $12$ AM). This indicates that being active at $6$ AM is much more predictive of being active at $7$ AM ($\rho\approx0.6$) than being active at $3$ AM for being active at $4$ AM ($\rho \approx 0.3$). These results are highly significant at all times of the day, the association is always positive, and the overall shape of the correlation is relatively stable across subgroups. 

The second panel (corresponding to lag $T=4$ hour correlation) indicates positive correlations throughout the day, except between $1$ AM and $3$ AM, when the correlation is slightly negative, but small ($\rho \approx -0.05$). This could be due to the fact that study participants who sleep better (e.g., less disturbed sleep) at $2$ AM are more likely to be active at $6$ AM. This is not unreasonable, but we will not over-interpret this small negative correlation. Relative to the first panel, correlations in the second panel decrease substantially in absolute value to a range of $0.05$ to $0.4$, or roughly half the correlations observed for lag $T=1$ hour. As before, the general shape of these associations is stable among the four groups with slightly smaller correlations for older study participants.

Panels 3 and 4 in Figure~\ref{fig:corlag-panels} provide the same information as panels $1$ and $2$, but for lag $T=8$ and $T=12$ hours, respectively. Results are intuitive, though it was not possible to quantify them before conducting this analysis. Correlations for lag $T=8$ hours vary between $-0.2$ and $0.2$, with positive correlations between $6$ AM and $5$ PM and negative correlations elsewhere. Correlations for lag $T=12$ hours show two periods, likely because of the diurnal/nocturnal associations of physical activity. The absolute value of the correlations tends to be much smaller, typically less than $0.1$. This is expected, as predicting $12$ hours ahead how active a person is is much harder than predicting $1$ hour ahead. As with the first two panels, results are quite consistent across subgroups. It is worth noting that the correlations have highly complex structures, get smaller with time lag $T$, but not in a uniform manner across the range, and never collapse to zero. This is likely due to our definition of time, which wraps around midnight. To the best of our knowledge, this is the first time these components of the variation of objectively measured physical activity in humans have been revealed and quantified.

\subsubsection{Modeling the skewness and excess kurtosis.}

\begin{figure}
    \centering
    \includegraphics[width=\linewidth]{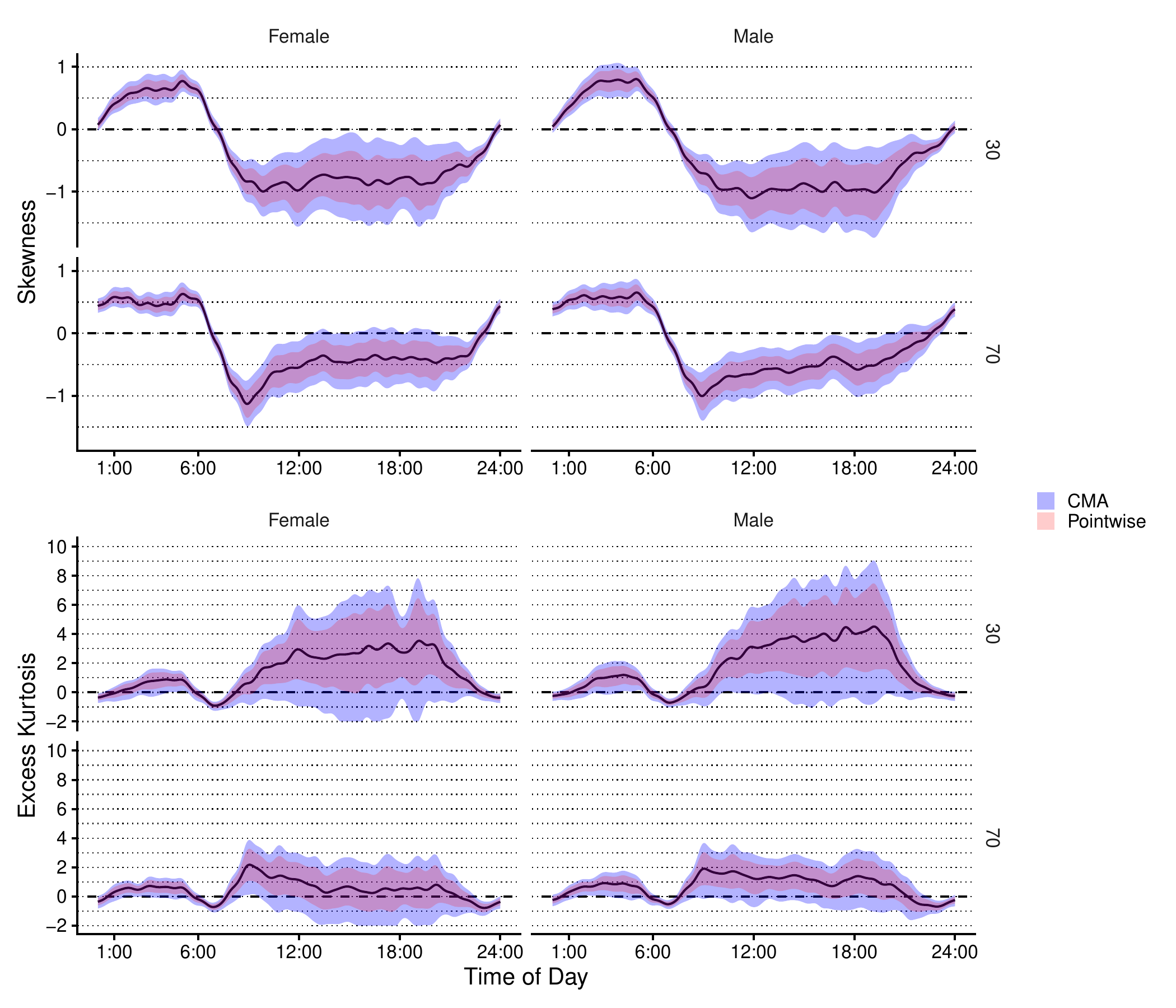}
    \caption{Model-based estimated skewness (top panel) and excess kurtosis (bottom panel) of log-MIMS outcome. Sub-panels correspond to the same subgroups as in Figure \ref{fig:resid-var}. Pointwise (blue) and asymmetric Correlation and Multiplicity Adjusted (CMA, red) 95\% confidence bands are included.}
    \label{fig:skewkurt_panel}
\end{figure}

Figure \ref{fig:skewkurt_panel} displays the model-based estimates of skewness  and excess kurtosis of the log-MIMS as a function of time of the day for the same subgroups as in Figure \ref{fig:resid-var}. 
Results indicate that after log transformation, the marginal distributions of residuals exhibit statistically significant positive skew during the night (12 AM - 6 AM) and negative skew during the day (6 AM - 10 PM). While the positive skew may be expected, the negative skew is not and indicates that, without careful review of the data one can easily over- or under-transform data. Here we argue for a ``trust but verify" or, more precisely, ``hope but plot", approach. On a positive note the skew is not very large, indicating that it may not substantially affect the inference on the fixed effects, though it may have a substantial effect on prediction intervals.

The excess kurtosis of the log-MIMS is not significantly different from zero during the daytime, but is negative (platykurtic) between 6 AM and 7 AM. 
This is intuitive, as it is common for people to wake up between 6 AM and 7 AM, which is consistent with many zeros (no activity) for people still sleeping and some larger measurements for people who have woken up, which can lead to distributions that have positive skew and insufficient kurtosis. To the best of our knowledge, this is the first time these behaviors of the data have been illustrated and quantified. 
While the mean, standard deviation, and kurtosis showed substantial differences between the age and gender groups, the correlation and skewness did not. It is important to note that point estimation, smoothing, and bootstrapping of study participants remain robust to departures from the GP assumption. More importantly, our methods can easily be implemented using traditional analytic techniques extended to functional data.

\section{Discussion}
\label{discussion}

We analyzed the NHANES minute-level activity dataset and estimated the conditional functional moments given subject characteristics such as age, BMI, and gender.
The activity data exhibit substantial deviations from the Gaussian distribution and show strong dependence on covariates, as indicated by our analysis of the functional skewness and excess kurtosis.  Applying a transformation could reduce non-Gaussianity in certain parts of the model, while having unintended consequences in others. One could consider different transformations at various points in the time domain, but this quickly becomes impractical and has substantial consequences for the translation of findings.

The method we have proposed can be applied using any basis, and it could be mapped back to the original scale of the data. However, choosing a spline basis in which the parameters have a local interpretation can substantially improve translation, as the visual mapping can be temporally localized. By working with the interpretable local moments, we provide a pragmatic and easy to implement methodology. Can these methods be extended to the full local distribution? Probably. Should they be in this paper? We do not believe so.

\section{Software}
\label{software}

Code for the simulations in Section \ref{sec:simulation} and the MIMS analysis in Section \ref{sec:results} is available at the GitHub repository https://github.com/LMY99/functional-moment-regression-code. 
\newpage
\section{Supplementary Material}
\label{supp}

\subsection{Explicit Formula of Conditional Kurtosis under the Model Setting in Section \ref{subsec:joint_model}}

Similar to the formula of conditional skewness, we can also derive formula for the conditional kurtosis. Note that for two independent random variables $X,Y$ with mean zero, $E[(X+Y)^4]=E[X^4]+E[Y^4]+6E[X^2]E[Y^2]$, so we have that
$$
\begin{aligned}
E[\{Y_i(s)-E[Y_i(s)|\mathbf{X}_i]\}^4|\mathbf{X}_i]&=E[\{\sum\limits_{j=1}^J\xi_{ij}\phi_j(s)-E[\sum\limits_{j=1}^J\xi_{ij}\phi_j(s)|\mathbf{X}_i]\}^4|\mathbf{X}_i]
\\&+E[\epsilon_i^4(s)|\mathbf{X}_i]
\\&+6\text{Var}[\epsilon_i(s)|\mathbf{X}_i]\text{Var}[\sum\limits_{j=1}^J\xi_{ij}\phi_j(s)|\mathbf{X_i}]
\end{aligned}
$$
Under the normal assumption of $\epsilon_i(s)$, $E[\epsilon_i^4(s)]=3\sigma^4_\epsilon(s)$. Thus, 
$$
\begin{aligned}
    &\text{Kurtosis}[Y_i(s)|\mathbf{X}_i]=\frac{K(s)+3\sigma_\epsilon^4(s)+6\sigma_\epsilon^2(s)\sum_{j_1,j_2=1}^J e^{\XX_i^t(\boldsymbol{\gamma}_{j_1}+\boldsymbol{\gamma}_{j_2})/2}\CC_{j_1,j_2}\phi_{j_1}(s)\phi_{j_2}(s)}
    {[\sum_{j_1,j_2=1}^J e^{\XX_i^t(\boldsymbol{\gamma}_{j_1}+\boldsymbol{\gamma}_{j_2})/2}\CC_{j_1,j_2}\phi_{j_1}(s)\phi_{j_2}(s)+\sigma^2_\epsilon(s)]^{2}}\;.
\\\text{where}\\
&K(s)=\sum\limits_{j_1,j_2,j_3,j_4=1}^J\phi_{j_1}(s)\phi_{j_2}(s)\phi_{j_3}(s)\phi_{j_4}(s)u(\mathbf{X}_i;j_1,j_2,j_3,j_4)e^{\mathbf{X}_i^t(\boldsymbol{\gamma}_{j_1}+\boldsymbol{\gamma}_{j_2}+\boldsymbol{\gamma}_{j_3}+\boldsymbol{\gamma}_{j_4})/2},\\
&u(\mathbf{X}_i;j_1,j_2,j_3,j_4)=\begin{cases}
e^{\mathbf{X}_i^t\boldsymbol{\eta}_{j_1,j_2,j_3,j_4}}&\sum\limits_{m=1}^4I(j_m=j)=0,2\text{ or } 4\text{ for all }j=1,2,\ldots,J\\
\mathbf{X}_i^t\boldsymbol{\eta}_{j_1,j_2,j_3,j_4}&\text{Otherwise}\;\end{cases}
\end{aligned}
$$
\newpage
\subsection{Simulation Parameters}

The fixed effect parameters were generated as:
\[
\begin{aligned}
\beta_0(s)&=2.5-1.8\exp\{2(1-\cos[(2s-\frac{5}{18})\pi])\}\\
\beta_1(s)&=\begin{cases}
0.005-0.01(1-\cos\frac{(3s-1)\pi}{2})&1/3\le s\le 1\\
0.005-0.01(1-\cos[(3s-1)\pi])&0\le s< 1/3
\end{cases}\\
\beta_2(s)&=\begin{cases}
0.1-0.125(1-\cos\frac{(s+1/6)\pi}{3/8})&0\le s<5/24\\
0.1-0.125(1-\cos\frac{(s-5/12)\pi}{5/24})&5/24\le s<5/12\\
0.1&5/12\le s<5/6\\
0.1-0.125(1-\cos\frac{(s-5/6)\pi}{29/24})&5/6\le s<1
\end{cases}\\
\beta_3(s)&=\begin{cases}
0.02-0.125(1-\cos\frac{(s-1/8)\pi}{5/6})&0\le s<1/8\\
0.02-0.125(1-\cos\frac{(s-1/8)\pi}{1/6})&1/8\le s<7/24\\
-0.23+0.125(1-\cos\frac{(s-7/24)\pi}{5/6})&7/24\le s<1
\end{cases}
\end{aligned}
\]

The random effect coefficients were generated as described in Algorithm \ref{alg:cap}.

\begin{algorithm}[htbp]
\caption{\\Generating random effect coefficients from a correlated non-Gaussian distribution}\label{alg:cap}
\begin{algorithmic}
\REQUIRE Covariates $(x_1,x_2,x_3)$
\STATE  1. Sample
    $(\gamma_1,\gamma_2,\gamma_3,\gamma_4,\gamma_5)\sim N(0,\Sigma_0)$
    where $$
    \Sigma_0=\begin{pmatrix}
    1& -0.4& 0.7& -0.4& -0.2 \\
    -0.4& 1& -0.4& 0.5& -0.1 \\
    0.7& -0.4& 1& -0.5& 0.1 \\
    -0.4& 0.5& -0.5& 1& -0.5\\
    -0.2& -0.1& 0.1& -0.5& 1\\
    \end{pmatrix}
    $$

\STATE  2. Let $$
    \begin{aligned}
    \xi_1^*&=-(F^{-1}(\Phi(-\gamma_1);3)-3)/\sqrt{3}\\
    \xi_2^*&=(F^{-1}(\Phi(\gamma_2);50)-50)/\sqrt{50}\\
    \xi_3^*&=-(F^{-1}(\Phi(-\gamma_3);20)-20)/\sqrt{20}\\
    \xi_4^*&=\gamma_4\\
    \xi_5^*&=(F^{-1}(\Phi(\gamma_5);4)-4)/\sqrt{4}\\
    \end{aligned}
    $$ where $\Phi$ is the CDF of $N(0,1)$, and $F^{-1}(q;\alpha)$ is
    the $q$-quantile of Gamma distribution with shape $\alpha$ and scale
    $1$.
\STATE 3. Let $\xi_j=\xi_j^*\exp\big[\frac{1}{2}(b_{0j}+\sum\limits_{i=1}^3b_{ij}x_i)\big]$ where $$
    B=(b_{ij})_{0\le i\le 3, 1\le j\le 5}=\begin{pmatrix}
    -0.2&-0.009&-0.04&-0.07\\
    -0.01&-0.05&-0.1&-0.06\\
    -0.03&0.2&0.1&-0.005\\
    0.01&-0.8&-0.006&-0.2\\
    -1&-0.003&0.02&-0.02
    \end{pmatrix}
    $$
\end{algorithmic}
\end{algorithm}

\begin{figure}
    \centering
    \includegraphics[width=0.8\linewidth, page=1]{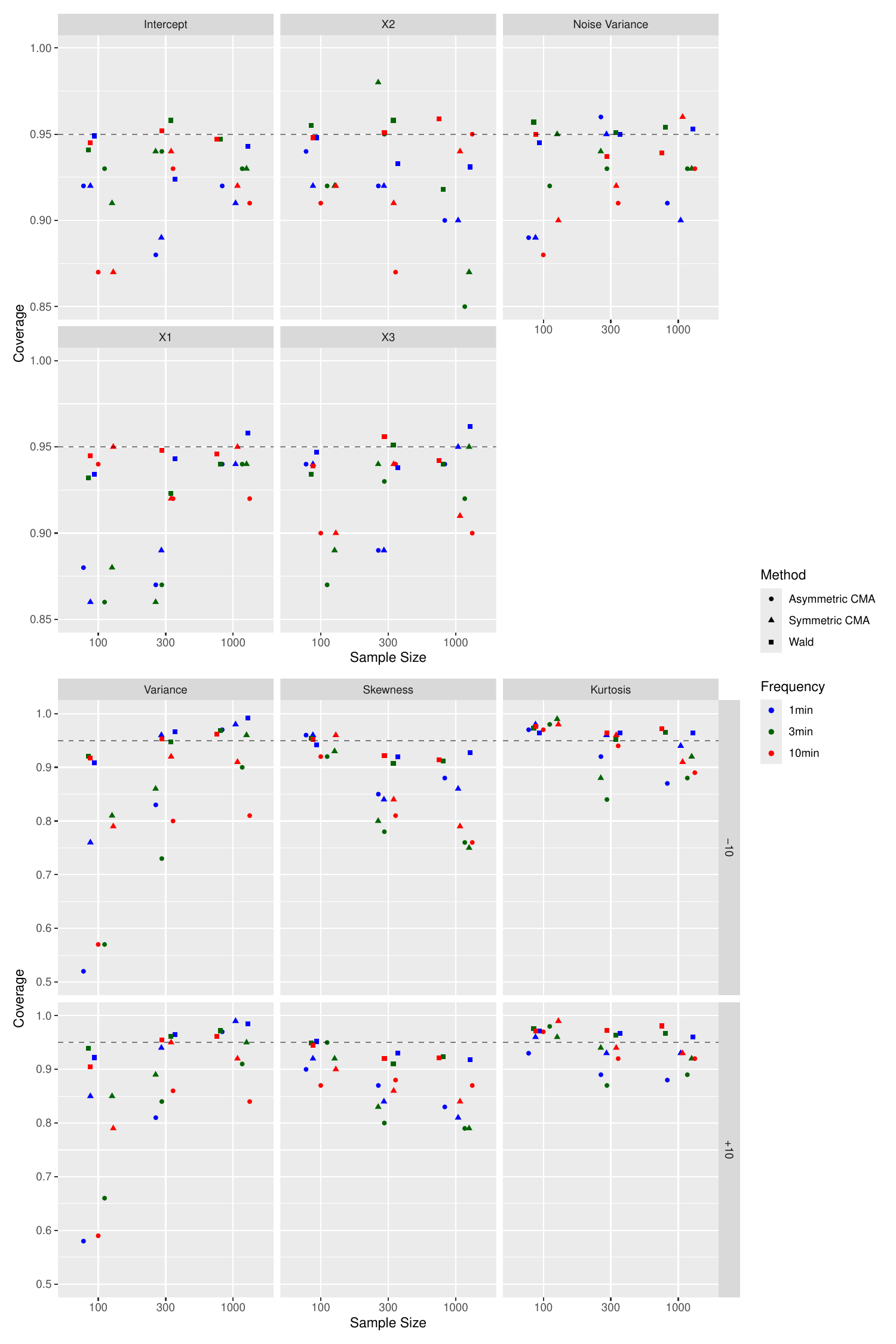}
    \caption{Scatterplot of simulated empirical coverage probability of multiple functional parameters, under different number of individuals (N) and different sampling frequencies. 
    The top panel includes, top to bottom, left to right, the intercept, three fixed effect parameters and noise variance.
    The bottom panel includes, left to right, the conditional variance, skewness and excess kurtosis, and the panels on the first and second rows indicate covariate values of $(-10,0,0)$ and $(10,0,0)$ respectively. 
    Global coverage probabilities are plotted for symmetric and asymmetric CMA intervals; average pointwise probabilities over the domain are plotted for Wald intervals.
    The points are jittered horizontally for clearer presentation.
    }
    \label{fig:coverage}
\end{figure}

\begin{table}[htbp]
	\centering
	\caption{Empirical coverage probability of confidence intervals for fixed effects and noise variance. 
		For symmetric and asymmetric Correlation and Multiplicity Adjusted (CMA) intervals, we show the global coverage probability;
		For Wald intervals, we show the coverage probability averaged over the entire curve. 
		Sample sizes are $n=100,300,1000$; frequencies are 1, 3, and 10 minutes.}
	\label{tab:coverage_FE_noise}
	\begin{tabular}{cccccc}
		\hline
		Method & Parameter & Frequency & N=100 & N=300 & N=1000 \\
		\hline
		Symmetric CMA & Intercept & 1  & 0.92 & 0.89 & 0.91 \\
		Symmetric CMA & Intercept & 3  & 0.91 & 0.94 & 0.93 \\
		Symmetric CMA & Intercept & 10 & 0.87 & 0.94 & 0.92 \\
		Asymmetric CMA & Intercept & 1  & 0.92 & 0.88 & 0.92 \\
		Asymmetric CMA & Intercept & 3  & 0.93 & 0.94 & 0.93 \\
		Asymmetric CMA & Intercept & 10 & 0.87 & 0.93 & 0.91 \\
		Wald & Intercept & 1  & 0.949 & 0.924 & 0.943 \\
		Wald & Intercept & 3  & 0.941 & 0.958 & 0.947 \\
		Wald & Intercept & 10 & 0.945 & 0.952 & 0.947 \\
		\hline
		Symmetric CMA & $X_1$ & 1  & 0.86 & 0.89 & 0.94 \\
		Symmetric CMA & $X_1$ & 3  & 0.88 & 0.86 & 0.94 \\
		Symmetric CMA & $X_1$ & 10 & 0.95 & 0.92 & 0.95 \\
		Asymmetric CMA & $X_1$ & 1  & 0.88 & 0.87 & 0.94 \\
		Asymmetric CMA & $X_1$ & 3  & 0.86 & 0.87 & 0.94 \\
		Asymmetric CMA & $X_1$ & 10 & 0.94 & 0.92 & 0.92 \\
		Wald & $X_1$ & 1  & 0.934 & 0.943 & 0.958 \\
		Wald & $X_1$ & 3  & 0.932 & 0.923 & 0.940 \\
		Wald & $X_1$ & 10 & 0.945 & 0.948 & 0.946 \\
		\hline
		Symmetric CMA & $X_2$ & 1  & 0.92 & 0.92 & 0.90 \\
		Symmetric CMA & $X_2$ & 3  & 0.92 & 0.98 & 0.87 \\
		Symmetric CMA & $X_2$ & 10 & 0.92 & 0.91 & 0.94 \\
		Asymmetric CMA & $X_2$ & 1  & 0.94 & 0.92 & 0.90 \\
		Asymmetric CMA & $X_2$ & 3  & 0.92 & 0.95 & 0.85 \\
		Asymmetric CMA & $X_2$ & 10 & 0.91 & 0.87 & 0.95 \\
		Wald & $X_2$ & 1  & 0.948 & 0.933 & 0.931 \\
		Wald & $X_2$ & 3  & 0.955 & 0.958 & 0.918 \\
		Wald & $X_2$ & 10 & 0.948 & 0.951 & 0.959 \\
		\hline
		Symmetric CMA & $X_3$ & 1  & 0.94 & 0.89 & 0.95 \\
		Symmetric CMA & $X_3$ & 3  & 0.89 & 0.94 & 0.95 \\
		Symmetric CMA & $X_3$ & 10 & 0.90 & 0.94 & 0.91 \\
		Asymmetric CMA & $X_3$ & 1  & 0.94 & 0.89 & 0.94 \\
		Asymmetric CMA & $X_3$ & 3  & 0.87 & 0.93 & 0.92 \\
		Asymmetric CMA & $X_3$ & 10 & 0.90 & 0.94 & 0.90 \\
		Wald & $X_3$ & 1  & 0.947 & 0.938 & 0.962 \\
		Wald & $X_3$ & 3  & 0.934 & 0.951 & 0.940 \\
		Wald & $X_3$ & 10 & 0.939 & 0.956 & 0.942 \\
		\hline
		Symmetric CMA & Noise Variance & 1  & 0.89 & 0.95 & 0.90 \\
		Symmetric CMA & Noise Variance & 3  & 0.95 & 0.94 & 0.93 \\
		Symmetric CMA & Noise Variance & 10 & 0.90 & 0.92 & 0.96 \\
		Asymmetric CMA & Noise Variance & 1  & 0.89 & 0.96 & 0.91 \\
		Asymmetric CMA & Noise Variance & 3  & 0.92 & 0.93 & 0.93 \\
		Asymmetric CMA & Noise Variance & 10 & 0.88 & 0.91 & 0.93 \\
		Wald & Noise Variance & 1  & 0.945 & 0.950 & 0.953 \\
		Wald & Noise Variance & 3  & 0.957 & 0.951 & 0.954 \\
		Wald & Noise Variance & 10 & 0.950 & 0.937 & 0.939 \\
	\end{tabular}
\end{table}

\begin{table}[htbp]
\footnotesize
	\centering
	\caption{Simulation results: coverage probability of confidence intervals for conditional variance/skewness/kurtosis given covariates. 
		Global coverage refers to simultaneous confidence bands (symmetric/asymmetric bandwidths), 
		while pointwise coverage refers to Wald intervals. 
		Sample sizes are $n=100,300,1000$; frequencies are 1, 3, and 10 minutes. 
		Results are reported for two covariate values of $X$, $X_1=(-10,0,0)$ and $X_2=(10,0,0)$.}
	\label{tab:coverage_cond_moments}
	
	\begin{tabular}{ccccccc}
		\hline
		Method & Parameter & Covariate & Frequency & N=100 & N=300 & N=1000 \\
		\hline
		Symmetric CMA & Variance & $(-10,0,0)$ & 1  & 0.76 & 0.96 & 0.98 \\
		Symmetric CMA & Variance & $(-10,0,0)$ & 3  & 0.81 & 0.86 & 0.96 \\
		Symmetric CMA & Variance & $(-10,0,0)$ & 10 & 0.79 & 0.92 & 0.91 \\
		Symmetric CMA & Variance & $(~~10,0,0)$ & 1  & 0.85 & 0.94 & 0.99 \\
		Symmetric CMA & Variance & $(~~10,0,0)$ & 3  & 0.85 & 0.89 & 0.95 \\
		Symmetric CMA & Variance & $(~~10,0,0)$ & 10 & 0.79 & 0.95 & 0.92 \\
		Asymmetric CMA & Variance & $(-10,0,0)$ & 1  & 0.52 & 0.83 & 0.97 \\
		Asymmetric CMA & Variance & $(-10,0,0)$ & 3  & 0.57 & 0.73 & 0.90 \\
		Asymmetric CMA & Variance & $(-10,0,0)$ & 10 & 0.57 & 0.80 & 0.81 \\
		Asymmetric CMA & Variance & $(~~10,0,0)$ & 1  & 0.58 & 0.81 & 0.97 \\
		Asymmetric CMA & Variance & $(~~10,0,0)$ & 3  & 0.66 & 0.84 & 0.91 \\
		Asymmetric CMA & Variance & $(~~10,0,0)$ & 10 & 0.59 & 0.86 & 0.84 \\
		Wald & Variance & $(-10,0,0)$ & 1  & 0.908 & 0.966 & 0.992 \\
		Wald & Variance & $(-10,0,0)$ & 3  & 0.921 & 0.947 & 0.969 \\
		Wald & Variance & $(-10,0,0)$ & 10 & 0.917 & 0.953 & 0.962 \\
		Wald & Variance & $(~~10,0,0)$ & 1  & 0.922 & 0.965 & 0.985 \\
		Wald & Variance & $(~~10,0,0)$ & 3  & 0.939 & 0.961 & 0.972 \\
		Wald & Variance & $(~~10,0,0)$ & 10 & 0.905 & 0.955 & 0.961 \\
		\hline
		Symmetric CMA & Skewness & $(-10,0,0)$ & 1  & 0.96 & 0.84 & 0.86 \\
		Symmetric CMA & Skewness & $(-10,0,0)$ & 3  & 0.93 & 0.80 & 0.75 \\
		Symmetric CMA & Skewness & $(-10,0,0)$ & 10 & 0.96 & 0.84 & 0.79 \\
		Symmetric CMA & Skewness & $(~~10,0,0)$ & 1  & 0.92 & 0.84 & 0.81 \\
		Symmetric CMA & Skewness & $(~~10,0,0)$ & 3  & 0.92 & 0.83 & 0.79 \\
		Symmetric CMA & Skewness & $(~~10,0,0)$ & 10 & 0.90 & 0.86 & 0.84 \\
		Asymmetric CMA & Skewness & $(-10,0,0)$ & 1  & 0.96 & 0.85 & 0.88 \\
		Asymmetric CMA & Skewness & $(-10,0,0)$ & 3  & 0.92 & 0.78 & 0.76 \\
		Asymmetric CMA & Skewness & $(-10,0,0)$ & 10 & 0.92 & 0.81 & 0.76 \\
		Asymmetric CMA & Skewness & $(~~10,0,0)$ & 1  & 0.90 & 0.87 & 0.83 \\
		Asymmetric CMA & Skewness & $(~~10,0,0)$ & 3  & 0.95 & 0.80 & 0.79 \\
		Asymmetric CMA & Skewness & $(~~10,0,0)$ & 10 & 0.87 & 0.88 & 0.87 \\
		Wald & Skewness & $(-10,0,0)$ & 1  & 0.942 & 0.920 & 0.927 \\
		Wald & Skewness & $(-10,0,0)$ & 3  & 0.954 & 0.907 & 0.912 \\
		Wald & Skewness & $(-10,0,0)$ & 10 & 0.952 & 0.922 & 0.914 \\
		Wald & Skewness & $(~~10,0,0)$ & 1  & 0.952 & 0.930 & 0.918 \\
		Wald & Skewness & $(~~10,0,0)$ & 3  & 0.949 & 0.910 & 0.924 \\
		Wald & Skewness & $(~~10,0,0)$ & 10 & 0.945 & 0.920 & 0.921 \\
		\hline
		Symmetric CMA & Kurtosis & $(-10,0,0)$ & 1  & 0.98 & 0.96 & 0.94 \\
		Symmetric CMA & Kurtosis & $(-10,0,0)$ & 3  & 0.99 & 0.88 & 0.92 \\
		Symmetric CMA & Kurtosis & $(-10,0,0)$ & 10 & 0.98 & 0.96 & 0.91 \\
		Symmetric CMA & Kurtosis & $(~~10,0,0)$ & 1  & 0.96 & 0.93 & 0.93 \\
		Symmetric CMA & Kurtosis & $(~~10,0,0)$ & 3  & 0.96 & 0.94 & 0.92 \\
		Symmetric CMA & Kurtosis & $(~~10,0,0)$ & 10 & 0.99 & 0.94 & 0.93 \\
		Asymmetric CMA & Kurtosis & $(-10,0,0)$ & 1  & 0.97 & 0.92 & 0.87 \\
		Asymmetric CMA & Kurtosis & $(-10,0,0)$ & 3  & 0.98 & 0.84 & 0.88 \\
		Asymmetric CMA & Kurtosis & $(-10,0,0)$ & 10 & 0.97 & 0.94 & 0.89 \\
		Asymmetric CMA & Kurtosis & $(~~10,0,0)$ & 1  & 0.93 & 0.89 & 0.88 \\
		Asymmetric CMA & Kurtosis & $(~~10,0,0)$ & 3  & 0.98 & 0.87 & 0.89 \\
		Asymmetric CMA & Kurtosis & $(~~10,0,0)$ & 10 & 0.97 & 0.92 & 0.92 \\
		Wald & Kurtosis & $(-10,0,0)$ & 1  & 0.964 & 0.964 & 0.964 \\
		Wald & Kurtosis & $(-10,0,0)$ & 3  & 0.973 & 0.952 & 0.965 \\
		Wald & Kurtosis & $(-10,0,0)$ & 10 & 0.976 & 0.964 & 0.972 \\
		Wald & Kurtosis & $(~~10,0,0)$ & 1  & 0.971 & 0.967 & 0.960 \\
		Wald & Kurtosis & $(~~10,0,0)$ & 3  & 0.976 & 0.964 & 0.967 \\
		Wald & Kurtosis & $(~~10,0,0)$ & 10 & 0.971 & 0.973 & 0.981 \\
	\end{tabular}
\end{table}
\newpage
\section*{Acknowledgments}

{\it Conflict of Interest}: None declared.

\bibliographystyle{biorefs}
\bibliography{refs}

\end{document}

%% file: tables/table_1_characteristics.tex
\begin{table}
\centering
\begin{tblr}[         
]                     
{                     
colspec={Q[]Q[]Q[]Q[]},
column{1,2}={}{halign=l,},
column{3,4}={}{halign=r,},
hline{5}={1,2,3,4}{solid, black, 0.05em},
}                     
\toprule
&    & \textbf{Mean} & \textbf{Std. Dev.} \\ \midrule 
Age (years) &  & \num{50.0} & \num{17.5} \\
Body Mass Index (BMI) (kg/m$^2$) &  & \num{29.2} & \num{7.1} \\
Poverty Income Ratio (PIR) &  & \num{2.5} & \num{1.6} \\
&  & \textbf{N} & $\mathbf{\mathbf{(\%)}}$ \\
Gender & Male & \num{3637} & \num{48.0} \\
& Female & \num{3941} & \num{52.0} \\
Race & Mexican American & \num{843} & \num{11.1} \\
& Other Hispanic & \num{717} & \num{9.5} \\
& Non-Hispanic White & \num{3195} & \num{42.2} \\
& Non-Hispanic Black & \num{1763} & \num{23.3} \\
& Non-Hispanic Asian & \num{837} & \num{11.0} \\
& Other Race & \num{223} & \num{2.9} \\
History of Coronary Heart Disease (CHD) & No & \num{7263} & \num{95.8} \\
& Yes & \num{315} & \num{4.2} \\
Education Level & Less than high school & \num{1613} & \num{21.3} \\
& High school equivalent & \num{1662} & \num{21.9} \\
& More than high school & \num{4303} & \num{56.8} \\
\bottomrule
\end{tblr}
\caption{Summary characteristics of the participants included in this analysis in terms of the covariates included in the model.}
\label{tab:tab1-char}
\end{table}